\documentclass[10pt,aps,pra,twocolumn,superscriptaddress,nofootinbib,longbibliography]{revtex4-2}
\usepackage[T1]{fontenc}
\usepackage[utf8]{inputenc}
\usepackage{lmodern}

\usepackage{amsmath,amssymb,amsthm,mathtools,bm}
\usepackage{physics}         
\usepackage{booktabs}          
\usepackage{graphicx}
\usepackage{float}
\usepackage{xcolor}
\usepackage{array}            
\usepackage{hyperref}
\hypersetup{colorlinks=true,linkcolor=blue,citecolor=blue,urlcolor=blue}

\newcolumntype{P}[1]{>{\raggedright\arraybackslash}p{#1}}

\begin{document}

\title{Quantum gravimetry with intrinsic quantum time uncertainty}

\author{Salman Sajad Wani}
\affiliation{Qatar Center for Quantum Computing, College of Science and Engineering, Hamad Bin Khalifa University, Doha, Qatar}

\author{Sundus Abdi}
\affiliation{Department of Mathematical and Computational Sciences, University of Toronto, Ontario, Canada}

\author{Rushda Naik}
\affiliation{Canadian Quantum Research Center, 460 Doyle Ave 106, Kelowna, BC V1Y 0C2, Canada}
\author{Saif Al-Kuwari}
\affiliation{Qatar Center for Quantum Computing, College of Science and Engineering, Hamad Bin Khalifa University, Doha, Qatar}
\begin{abstract}
 We study quantum gravimetry when the interrogation time carries intrinsic uncertainty, motivated by a fundamental limit on temporal resolution associated with the energy--time uncertainty relation. For linearly gravity-coupled gravimeters, we obtain the effective gravity information by profiling the interrogation time from the two-parameter quantum Fisher information (QFI) matrix. In this class, the time-information block is quadratic in the gravitational parameter, and for quadratic background dynamics, the gravity--time cross term becomes affine in $g$. These properties yield a normalized expression for the fraction of standard single-parameter gravity QFI that remains once interrogation time is treated as a nuisance parameter, with an affine numerator and a Lorentzian denominator. We work out these results in three benchmark models: a freely falling Gaussian wavepacket, the Kasevich--Chu light-pulse atom interferometer, and an idealized closed-unitary optomechanical model. The Gaussian free-fall benchmark yields an exact closed-form expression for the effective gravity information and shows explicitly how nuisance-time profiling suppresses the momentum-spread-dependent part of the standard single-parameter gravity QFI. In the Kasevich--Chu interferometer, internal state population readout gives a rank-deficient measured two-parameter geometry unless independent timing information is supplied, whereas full access to the final motional and internal states restores a full-rank geometry with retention controlled by the competition between initial velocity spread and gravitationally accumulated motion. In atom-interferometric benchmarks, the framework yields explicit conditions for minimizing nuisance-time information loss, together with corresponding constraints on momentum spread, spatial localization, and long-interrogation-time operation.
\end{abstract}
\maketitle

%=====================================================================
\section{Introduction}
\label{sec:intro}
%=====================================================================

Representative platforms for quantum gravimetry include atom-interferometric gravimeters based on freely falling atomic matter waves and optomechanical proposals to measure gravitational acceleration \cite{DeAngelis2009,Peters1999Nature,Menoret2018AQG,Armata2017OptomechGravity,Qvarfort2018Opto}. Related cold-atom sensor concepts are also being developed for ultralight-dark matter searches, long-baseline gravitational-wave detection, and field-deployable quantum sensing \cite{Bongs2019NatRevPhys,MAGIS1002021,AEDGE2020}. In light-pulse atom-interferometric phase-accumulation models, the gravity signal is encoded through a phase shift proportional to $gT^2$ \cite{DeAngelis2009,KasevichChu1992}. In the corresponding ideal single-parameter description, this gives a $T^4$ scaling of the Fisher information for $g$, with the interrogation time treated as a perfectly known classical control parameter.

This paper studies a benchmark estimation problem in which the encoded probe family depends jointly on gravitational acceleration and interrogation time. Following the nuisance-parameter formulation developed for generic linear quantum sensors \cite{Wani2026TimeUncertainty,Song2025TemporalResolution,HerbDegen2024QSL}, we treat the interrogation time as an encoded parameter of the state family and ask how much information about $g$ survives after the time direction is profiled out. The analysis concerns the information contained in the retained probe record. Technical noise sources such as timing jitter, synchronization error, and pulse noise are beyond the scope of this paper. We focus on pure-state unitary benchmark models in one spatial dimension, with gravity entering linearly through the sensing Hamiltonian. 

The central issue is identifiability: once gravity and interrogation time are placed on equal footing, what fraction of the nominal single-parameter gravity sensitivity remains accessible? For this class of sensors, the nuisance-time penalty acquires a constrained analytic form. The effective gravity information, defined later as the Schur complement of the time block of the two-parameter quantum Fisher information matrix, has a timing block that is quadratic in the gravitational parameter. Under a weak commutator condition on the $g$-independent dynamics, the gravity--time cross term is affine in $g$. Together, these properties yield a normalized retention law with an affine numerator and a Lorentzian denominator. While previous work developed the general framework for parameter nuisance for temporally uncertain sensing models \cite{Wani2026TimeUncertainty,Song2025TemporalResolution,HerbDegen2024QSL}, this paper identifies the closed structural form taken by this penalty within the class of linearly gravity-coupled unitary gravimeters and derives its consequences for representative gravimetric models.

We study three examples. The first is a freely falling Gaussian wavepacket, for which the full two-parameter quantum Fisher information matrix can be obtained exactly. The second is the Kasevich--Chu light-pulse atom interferometer \cite{KasevichChu1991,KasevichChu1992}. In the nuisance-parameter formalism used here, the relevant gravimetric figure of merit is the effective Fisher information obtained after profiling out the interrogation time, i.e., the Schur complement of the time block of the two-parameter quantum Fisher information matrix \cite{Wani2026TimeUncertainty,Szczykulska2016}. The Kasevich--Chu example shows how this quantity depends on the retained readout: internal state population measurements depend on $g$ and $T$ only through a single phase combination and therefore yield a rank-deficient measured geometry unless independent timing information is supplied, whereas access to both internal and motional degrees of freedom under ideal interferometric closure contributes a positive timing term and restores full rank. Any information about the duration of the interrogation beyond that contained in the retained probe record must then be entered explicitly, for example, through prior information \cite{Wani2026TimeUncertainty,Rubio2020Bayesian}. The third example is an idealized closed-unitary optomechanical model.

Within the full-state Kasevich--Chu benchmark, retaining substantial gravity information at long interrogation times requires a large initial longitudinal momentum variance. For pure states, this requirement is equivalent to tight initial longitudinal localization and carries the corresponding free-expansion and Doppler costs during subsequent evolution, consistent with the known sensitivity of long-$T$ atom interferometers to velocity spread, ballistic expansion, and contrast loss \cite{Dickerson2013PSI,Woltmann2024Collimation}. We quantify these trade-offs and compare them with source and interrogation-time scales reported for current atom gravimeters. The closed-system optomechanical limit serves as a geometric baseline \cite{Wani2026TimeUncertainty,Armata2017OptomechGravity,Qvarfort2018Opto}. In the optomechanical benchmark of Ref.~\cite{Qvarfort2018Opto}, homodyne detection is optimal at the decoupling times $t=2\pi n$, where the optical and mechanical sectors separate and the classical Fisher information reaches the quantum limit. The periodic revivals discussed here, therefore, describe the bare unitary correlation structure of the sensor prior to any open-system estimation, feedback, or continuously monitored readout layer. A full nuisance-time treatment for stochastic master-equation dynamics lies outside the present scope.

The paper is organized as follows. Section~\ref{sec:formalism} sets up the two-parameter estimation framework. Section~\ref{sec:kernel_condition_free} derives the structural kernel for linearly gravity-coupled sensors. Section~\ref{sec:continuous_free_fall} presents the exact free-fall Gaussian example. Section~\ref{sec:discrete_KC} analyzes the Kasevich--Chu interferometer under internal-only and full-state readout. Section~\ref{sec:universal_scaling} recasts the free-fall, interferometric, and optomechanical cases in a common normalized form. Section~\ref{sec:experimental_implications} quantifies the corresponding momentum variance and localization requirements for representative atom-gravimeter regimes.

\section{Multiparameter estimation with nuisance interrogation time}
\label{sec:formalism}

To rigorously quantify the parameter degeneracy that arises when interrogation time is included among the encoded parameters and treated as a nuisance parameter, we adopt the local-generator formulation of multiparameter quantum estimation \cite{Szczykulska2016,Albarelli2020Perspective}. 
We consider a time-independent linear sensor Hamiltonian of the form
\begin{equation}
\hat H(g)=\hat H_{0}+\lambda\, g\,\hat Q,
\label{eq:H_generic_linear}
\end{equation}
where $\hat H_{0}$ governs the $g$-independent background dynamics, $\hat Q$ is a Hermitian sensing observable, and $\lambda$ is the coupling strength. For a one-dimensional gravimeter based on the center-of-mass coordinate, one has $\lambda=m$ and $\hat Q=\hat z$. Starting from a normalized pure probe state $\ket{\psi_0}$, the encoding over an interrogation time $t$ is $\ket{\psi(g,t)}=\hat U(g,t)\ket{\psi_0},$ and $\hat U(g,t)=\exp\!\left(-\frac{i}{\hbar}\hat H(g)\,t\right).$ Because both gravity and the interrogation time shape the encoded state, we treat the family $\{\ket{\psi(g,t)}\}$ as a two-parameter estimation problem with parameter vector $(g,t)$. The sensitivity of the encoded family to parameter translations is captured by the pulled-back Hermitian local generators $ \hat{\mathcal H}_j(g,t):=
i\,\hat U^\dagger(g,t)\,\partial_j\hat U(g,t),$ with $ j\in\{g,t\}. $ Differentiation with respect to time gives the normalized energy generator $\hat{\mathcal H}_t(g,t)={\hat H(g)}/{\hbar}.$ For the gravity parameter, the Hadamard--Duhamel identity yields the exact integral representation \cite{PangBrun2014}.
\begin{equation}
\hat{\mathcal H}_g(g,t)
=
\frac{1}{\hbar}\int_0^t ds\,
\big(\partial_g \hat H(g)\big)_{H}(s),
\label{eq:Hg_general}
\end{equation}
where $(\cdot)_H(s)$ denotes Heisenberg evolution under $\hat H(g)$. For the linear family \eqref{eq:H_generic_linear}, this becomes
\begin{equation}
\hat{\mathcal H}_g(g,t)
=
\frac{\lambda}{\hbar}\int_0^t ds\,\hat Q_H(s).
\label{eq:Hg_linear}
\end{equation}
Thus, the time generator is set by the instantaneous Hamiltonian, while the gravity generator accumulates the Heisenberg-evolved sensing observable over the full interrogation window.

Throughout this section, we remain within the pure-state, unitary-encoding setting. For such families, the quantum Fisher information matrix (QFIM) is given by the symmetrized covariances of the local generators evaluated in the initial probe state \cite{Liu2020QFIMReview}:
\begin{equation}
F_{ij}(g,t)
=
4\,\operatorname{Cov}^{\mathrm{sym}}_{\psi_0}\!\big(\hat{\mathcal H}_i,\hat{\mathcal H}_j\big),
\end{equation}
where
$\operatorname{Cov}^{\mathrm{sym}}_{\psi_0}(A,B)
:=\frac{1}{2}\,\langle\{\Delta A,\Delta B\}\rangle_{\psi_0}$,
$\Delta A:=A-\langle A\rangle_{\psi_0}$,
$\langle\cdot\rangle_{\psi_0}:=\bra{\psi_0}\cdot\ket{\psi_0}$,
and
$\operatorname{Var}_{\psi_0}(A):=\operatorname{Cov}^{\mathrm{sym}}_{\psi_0}(A,A)$. In particular,

\begin{align}\label{eq:QFIM_entries}
  F_{gg} &= 4\,\operatorname{Var}_{\psi_0}(\hat{\mathcal{H}}_g),
    \notag\\
  F_{gt} &= 4\,\operatorname{Cov}^{\mathrm{sym}}_{\psi_0}
             (\hat{\mathcal{H}}_g,\hat{\mathcal{H}}_t),
    \notag\\
  F_{tt} &= 4\,\operatorname{Var}_{\psi_0}(\hat{\mathcal{H}}_t).
\end{align}

The diagonal entries $F_{gg}$ and $F_{tt}$ quantify the idealized single-parameter information for gravity and time, while the off-diagonal entry $F_{gt}$ measures the geometric correlation between the accumulation of gravitational and temporal phases. In the present formulation of nuisance parameters, $t$ acts as a quantum nuisance parameter: it strongly affects the output state, though it is not the parameter of interest. The quantity of information entering the inverse-QFIM bound for gravity after elimination of the timing coordinate is then the Schur complement of the $tt$ block, 
\begin{equation}
F_{\mathrm{eff}}(g,t)
:=
F_{gg}(g,t)-\frac{F_{gt}(g,t)^2}{F_{tt}(g,t)},
\qquad
F_{tt}(g,t)>0.
\label{eq:Feff_bare}
\end{equation}
For $N$ independent repetitions, the corresponding local multiparameter Cram\'er--Rao bound for locally unbiased estimators reads
$ \operatorname{Var}(\hat g)\ge {1}/{N\,F_{\mathrm{eff}}(g,t)}.$ Equation \eqref{eq:Feff_bare} is the central object for the nuisance-time structural analysis developed below: it quantifies how much of the nominal single-parameter gravity information survives after timing correlations are profiled out. Section~\ref{sec:kernel_condition_free} analyzes this irregular quantity for linearly gravity-coupled sensors and derives its universal structural form.

An additional layer arises when the timing sector is singular or when the estimation problem is supplemented by independent information about the evolution duration. In that setting, we combine the quantum Fisher information with an independent prior or auxiliary measurement that supplies Fisher information $ I^{(prior)}_t$ for the interrogation time. For a deterministic parameter $g$ and a nuisance parameter $t$ with independent prior information, the effective information for $g$ becomes
\begin{equation}
F_{\mathrm{eff}}^{(\mathrm{reg})}(g,t)
=
F_{gg}(g,t)-\frac{F_{gt}(g,t)^2}{F_{tt}(g,t)+I_t^{(\mathrm{prior})}},
\label{eq:Feff_schur_regularized}
\end{equation}
which follows from the additivity of Fisher information for independent data streams \cite{VanTrees1968,Rubio2020Bayesian}. In the interferometric setting discussed later, this regularization has a direct statistical meaning: the timing prior represents independent information about the evolution duration beyond that contained in the retained probe record. Quantum speed-limit considerations then provide a natural scale for the associated timing precision, as we quantify in Sec.~\ref{sec:experimental_implications} \cite{MargolusLevitin1998,HerbDegen2024QSL}.
However, at the level of the present section, this additional timing information is incorporated only through the additive information term $I_t^{(\mathrm{prior})}$; its physical implementation is left to the later architecture-specific discussion. This regularized form becomes physically necessary only in architectures for which the measured timing sector is singular, as in the phase-only internal readout of the Kasevich--Chu interferometer.
For the present nuisance-time problem, the Schur-complement quantity in Eq.~\eqref{eq:Feff_bare} is the appropriate quantum limit to estimate $g$ after eliminating the timing coordinate. In Ref.~\cite{Wani2026TimeUncertainty}, the same effective information arises when the interrogation time is treated explicitly as a nuisance parameter. In the complete nuisance-parameter formalism, Eq.~\eqref{eq:Feff_bare} coincides with the partial‑SLD Fisher information. Under the conditions established in~\cite{Suzuki2020} (Theorem~5.3), an optimal measurement and estimator can achieve the associated local unbiased bound. The SLD‑compatibility condition, which concerns the simultaneous saturation of the full multiparameter bound, is not required to optimize a single parameter in the presence of nuisance parameters \cite{Wani2026TimeUncertainty,Suzuki2020,SidhuKok2020}. Our focus here is the structural geometry of the nuisance-time penalty and its realization in representative gravimetric models.

\section{Universal structural kernel for linear gravimeters}
\label{sec:kernel_condition_free}

When interrogation time is treated as an encoded nuisance parameter without additional timing information, the gravity--time correlation generated by the encoded state itself dictates the nuisance-time penalty.
In this section, we derive the structural form of that penalty for linearly gravity-coupled one-dimensional sensors. The result has two layers: the timing block is always quadratic in the gravitational parameter, and under a weak commutator condition in the background dynamics, the gravity block $F_{gg}$ becomes independent of $g$ while the cross term $F_{gt}$ is affine in $g$, yielding a universal retention kernel. We specialize in
\begin{equation}
\hat H(g)=\hat H_0+m g\,\hat z,
\label{eq:H_linear_gravity}
\end{equation}
for which the first stage of the result is completely general within this Hamiltonian class, while the second stage uses the weak commutator condition 
and yields the full normalized form. The starting point is the time generator. We define the $g$-independent operators
$\hat A:={\hat H_0}/{\hbar}$, and $\hat B:={m\hat z}/{\hbar}.$ Thus, the time generator has the exact affine form $\hat{\mathcal H}_t(g,t)=\hat A+g\,\hat B. $ Substituting $\hat{\mathcal H}_t(g,t)$ into the covariance formula for the QFIM gives the following.
\begin{equation}
F_{tt}(g,t)
=
4\,\operatorname{Var}_{\psi_0}(\hat A+g\hat B)
=
c_0+c_1\,g+c_2\,g^2,
\label{eq:Ftt_quadratic_expand}
\end{equation}
with $c_0:=4\,\operatorname{Var}_{\psi_0}(\hat A)$, $c_1:=8\,\operatorname{Cov}^{\mathrm{sym}}_{\psi_0}(\hat A,\hat B),$ $c_2:=4\,\operatorname{Var}_{\psi_0}(\hat B).$ Accordingly, the nuisance-time block is always quadratic in the gravitational parameter. We assume throughout that the relevant second moments and symmetrized covariances in the initial state are finite. In the nondegenerate case $\operatorname{Var}_{\psi_0}(\hat z)>0$, so that $c_2>0$, completing the square yields
\begin{equation}
F_{tt}(g,t)
=
c_2\Big[(g-g_c)^2+g_*^2\Big],
\label{eq:Ftt_complete_square}
\end{equation}
where $g_c:=-\frac{c_1}{2c_2},$ and $ g_*^2:=\frac{c_0 c_2-c_1^2/4}{c_2^2}.$ The Cauchy--Schwarz inequality for the bilinear form of symmetric coefficients ensures $c_0 c_2-c_1^2/4\ge 0$, hence $g_*^2\ge 0$.
In the nondegenerate regime $g_*>0$, the inverse nuisance block factors through a Lorentzian kernel,
\begin{equation}
\frac{1}{F_{tt}(g,t)}
=
\frac{1}{c_2\,g_*^2}\,
\frac{1}{1+u^2},
\qquad
u:=\frac{g-g_c}{g_*}.
\label{eq:invFtt_lorentzian}
\end{equation}
The exceptional case $g_*=0$ corresponds to a degenerate timing block for which this normalized coordinate is undefined; therefore, the normalized kernel derived below is understood in the domain $g_*>0$.
At this stage, the geometry already reveals a universal denominator. The remaining question is the $g$-dependence of the gravity block and the cross term. To answer that question, we impose a weak commutator condition
on the background dynamics. Let $\hat z_0(s):=
e^{+\frac{i}{\hbar}\hat H_0 s}\,\hat z\,e^{-\frac{i}{\hbar}\hat H_0 s}$ denote the position operator evolved under $\hat H_0$. We assume that for all $\tau_1,\tau_2\in[0,t]$,
\begin{equation}
\big[\hat z_0(\tau_1),\hat z_0(\tau_2)\big]
=
i\hbar\,\Delta(\tau_1,\tau_2)\,\mathbb I,
\qquad
\Delta(\tau_1,\tau_2)\in\mathbb R.
\label{eq:weak_commutator_criterion}
\end{equation}
This condition is explicitly realized by the free-particle benchmark treated in Sec.~\ref{sec:continuous_free_fall} and by the closed-unitary optomechanical benchmark summarized in Appendix~\ref{app:opto_cross_term}. The general affine consequence for the gravity--time cross term is proved in Appendix~\ref{app:dyson_proof}. Under \eqref{eq:weak_commutator_criterion}, Appendix~\ref{app:dyson_proof} shows that the interaction-picture Magnus expansion truncates exactly, yielding an affine scalar shift in the Heisenberg position. Reference~\cite{Blanes2009Magnus} provides a general background on the Magnus expansion.
As a result, the interacting position operator takes the form
\begin{equation}
\hat z_H(s;g)=\hat z_0(s)+g\,f(s)\,\mathbb I,
\label{eq:zH_affine}
\end{equation} for some real scalar function $f(s)$. Substituting \eqref{eq:zH_affine} into the generator formula \eqref{eq:Hg_linear} yields $\hat{\mathcal H}_g(g,t)
=\hat{\mathcal H}_g(0,t)+g\,\kappa(t)\,\mathbb I$, where $\hat{\mathcal H}_g(0,t)=\frac{m}{\hbar}\int_0^t ds\,\hat z_0(s),$ and $\kappa(t)=\frac{m}{\hbar}\int_0^t ds\,f(s).$ This structural truncation imposes two direct consequences on the Fisher geometry, because scalar identity shifts leave variances unchanged and the identity has zero symmetrized covariance with any operator. First, because the variance of an operator is invariant under a scalar identity shift, $F_{gg}(g,t) =4\,\operatorname{Var}_{\psi_0}\!\big(\hat{\mathcal H}_g(0,t)\big)
\equiv F_{gg}(t)$, so the gravity block is independent of $g$. Second, because the symmetrized covariance with the identity vanishes, the cross term reduces to
\begin{equation}
F_{gt}(g,t)
=
4\,\operatorname{Cov}^{\mathrm{sym}}_{\psi_0}\!\big(\hat{\mathcal H}_g(0,t),\hat A+g\hat B\big)
=
d_0(t)+d_1(t)\,g,
\label{eq:Fgt_affine}
\end{equation}
with $d_0(t):=
4\,\operatorname{Cov}^{\mathrm{sym}}_{\psi_0}\!\big(\hat{\mathcal H}_g(0,t),\hat A\big),$ and $d_1(t):=
4\,\operatorname{Cov}^{\mathrm{sym}}_{\psi_0}\!\big(\hat{\mathcal H}_g(0,t),\hat B\big).$ Thus, together with the unconditionally quadratic timing block, the weak commutator condition rigidly constrains the remaining geometry: the gravity block is independent of $g$, and the cross term is affine in $g$. The unregularized profiled gravity information therefore takes the form in Eq.~\eqref{eq:Feff_bare}. Introducing the normalized coordinate $u=(g-g_c)/g_*$ in the domain $g_*>0$, the quadratic timing element becomes $F_{tt}(g,t)=c_2 g_*^2(1+u^2)$.
Writing the affine cross term in normalized form,
\begin{equation}
F_{gt}(g,t)
=
\sqrt{F_{gg}(t)c_2}\,g_*
\big(\alpha_0(t)+\alpha_1(t)\,u\big),
\label{eq:Fgt_normalized}
\end{equation}
with $\alpha_0(t):=
{d_0(t)+d_1(t)g_c}/{\sqrt{F_{gg}(t)c_2}\,g_*},$ and $\alpha_1(t):=
{d_1(t)}/{\sqrt{F_{gg}(t)c_2}}$, we obtain the normalized master form $F_{\mathrm{eff}}(g,t)= F_{gg}(t)\,R(u;t)$, where
\begin{equation}
R(u;t)
=
1-\frac{\big(\alpha_0(t)+\alpha_1(t)\,u\big)^2}{1+u^2}.
\label{eq:R_master_universal}
\end{equation}
The coefficients $\alpha_0(t)$ and $\alpha_1(t)$ encode the model-dependent numerator structure, 
whereas the Lorentzian denominator has already been fixed by the quadratic structure of the timing block. This is the structural kernel underlying the later benchmarks. The nuisance-time penalty is carried by a Lorentzian denominator together with the square of an affine term in the numerator,
while the dependence of the platform enters only through the baseline scale $F_{gg}(t)$, the parameters of the axis $(g_c,g_*)$ and the dimensionless coefficients $(\alpha_0(t),\alpha_1(t))$.
Finally, the positive semidefiniteness of the QFIM gives
$F_{gt}(g,t)^2\le F_{gg}(t)\,F_{tt}(g,t)$, which in the normalized representation is equivalent to $\big(\alpha_0(t)+\alpha_1(t)\,u\big)^2\le 1+u^2$. Therefore, the retention factor obeys $0\le R(u;t)\le 1$ throughout the domain of physically admissible parameters. The free-fall benchmark of Sec.~\ref{sec:continuous_free_fall}, supported by Appendix~\ref{app:bch_algebra}, and the closed-unitary optomechanical benchmark of Appendix~\ref{app:opto_cross_term} realize this unregularized structural result directly.
By contrast, the phase-only Kasevich--Chu readout analyzed in Sec.~\ref{sec:discrete_KC}, with operator details given in Appendix~\ref{app:kc_operator_algebra}, requires the regularized extension discussed earlier because its measured timing sector is singular.

\section{Continuous metrology: free-falling Gaussian wavepackets}
\label{sec:continuous_free_fall}

As a first exact benchmark for the structural kernel derived in Sec.~\ref{sec:kernel_condition_free}, we consider a freely falling one-dimensional wavepacket in a uniform gravitational field, $\hat H(g)={\hat p^2}/{2m}+mg\,\hat z.$ The probe is prepared in a pure, centered, minimum-uncertainty Gaussian state, where centered means $\langle \hat z\rangle_{\psi_0}=\langle \hat p\rangle_{\psi_0}=0$,
with

\label{eq:Gaussian_moments}
\begin{align}
  \operatorname{Var}_{\psi_0}(\hat{z})
    &= \frac{\sigma^2}{2}, \qquad
    \operatorname{Var}_{\psi_0}(\hat{p})= \frac{\hbar^2}{2\sigma^2},
    \notag\\
  &\operatorname{Cov}^{\mathrm{sym}}_{\psi_0}(\hat{z},\hat{p})
    = 0.
\end{align}

In this model the background Hamiltonian is $\hat H_0=\hat p^2/(2m)$, so the free Heisenberg position is $\hat z_0(s)=\hat z+{\hat p}/{m}s.$ Hence,
\begin{equation}
[\hat z_0(u),\hat z_0(v)]
=
\left[\hat z+\frac{u}{m}\hat p,\hat z+\frac{v}{m}\hat p\right]
=
i\hbar\,\frac{v-u}{m}\,\mathbb I,
\label{eq:freefall_weak_condition}
\end{equation}
so the weak commutator criterion is exactly satisfied. The free-fall Gaussian therefore lies within the universal retention class described in Sec.~\ref{sec:kernel_condition_free}. For finite $\sigma$, the scale $g_*$ defined explicitly in Eq.~\eqref{eq:Sg_freefall} is strictly positive, so the nondegenerate normalized-kernel regime of Sec.~\ref{sec:kernel_condition_free} applies without further qualification. An exact Baker--Campbell-Hausdorff factorization of the unitary evolution yields closed forms for the local generators; the operator derivation is given in the Appendix~\ref{app:bch_algebra}. Evaluating the corresponding symmetrized covariances in the initial Gaussian state gives the exact unregularized QFIM for the parameter pair $(g,t)$. In the ordered basis $(g,t)$,
\begin{equation}
\bm F(g,t)=
\begin{pmatrix}
\displaystyle \frac{t^4}{2\sigma^2}+\frac{2m^2\sigma^2 t^2}{\hbar^2}
&
\displaystyle \frac{2m^2 g\sigma^2 t}{\hbar^2}
\\
\displaystyle \frac{2m^2 g\sigma^2 t}{\hbar^2}
&
\displaystyle \frac{\hbar^2}{2m^2\sigma^4}+\frac{2m^2 g^2\sigma^2}{\hbar^2}
\end{pmatrix}.
\label{eq:QFIM_Gaussian}
\end{equation}
This benchmark realizes the structural features of Sec.~\ref{sec:kernel_condition_free} in explicit form: the gravity block $F_{gg}$ is independent of $g$, the cross term $F_{gt}$ is affine in $g$ and in this case purely linear, and the nuisance-time block $F_{tt}$ is quadratic in $g$. Profiling out the nuisance time by the unregularized Schur complement yields
\begin{equation}
F_{\mathrm{eff}}(g,t)
=
\frac{t^4}{2\sigma^2}
+\frac{2m^2\sigma^2 t^2}{\hbar^2}\,S(g),
\label{eq:Feff_freefall}
\end{equation}
where
\begin{equation}
S(g)=\frac{1}{1+(g/g_*)^2},
\qquad
g_*=\frac{\hbar^2}{2m^2\sigma^3}.
\label{eq:Sg_freefall}
\end{equation}
Equation~\eqref{eq:Feff_freefall} makes the free-fall identifiability structure transparent. The profiled gravity information contains two distinct channels: an exactly retained $t^4$ contribution and a $t^2$ contribution multiplied by the Lorentzian factor $S(g)$. The nuisance-time penalty acts only on the latter. The term $t^4$ appears unsuppressed exactly in Eq.~\eqref{eq:Feff_freefall}; for fixed nonzero $g$ and finite $\sigma$, it dominates asymptotically at a large interrogation time. More specifically, for fixed nonzero $g$ and fixed $\sigma$, the large behavior $t$ satisfies
\begin{subequations}
\label{eq:rho_scaling_freefall}
\begin{align}
  F_{gg} &\sim t^4,   &  F_{gt} &\sim t,   \label{eq:rho_scaling_freefall}\\
  F_{tt} &\sim t^0,   &
  \rho^2 &:= \frac{F_{gt}^2}{F_{gg}F_{tt}} \sim t^{-2}.
  \label{eq:rho_scaling_freefall}
\end{align}
\end{subequations}
Thus, the gravity--time correlation decreases at long interrogation times. In this continuously evolving model, the momentum-variance-dependent part of $F_{gg}$, namely the term $2m^2\sigma^2 t^2/\hbar^2$ in Eq.~\eqref{eq:QFIM_Gaussian}, provides a contribution that, together with the cross term, enables the Schur complement to retain a contribution $t^4$ even after profiling over time. Under free evolution, that initial momentum variance manifests itself as wavepacket spreading, so the model can distinguish gravitational acceleration from interrogation time, while the exact unsuppressed $t^4$ contribution in Eq.~\eqref{eq:Feff_freefall} shows that this timing channel survives nuisance-time profiling.\\

The delocalized plane-wave limit corresponds to $\sigma\to\infty$, which enforces $\operatorname{Var}_{\psi_0}(\hat p)\to 0.$ This removes the initial momentum variance responsible for wavepacket expansion, thereby eliminating the kinematic degree of freedom that would otherwise allow the output state to encode information distinguishing gravity from interrogation time. Analytically, the geometric scale satisfies $g_*=\frac{\hbar^2}{2m^2\sigma^3}\propto \sigma^{-3},$ so for fixed nonzero $g$ the Lorentzian factor obeys  $S(g)\sim {g_*^2}/{g^2}\propto \sigma^{-6}.$  Hence the $t^2$ channel is completely suppressed, while the coefficient of the retained $t^4$ term in Eq.~\eqref{eq:Feff_freefall} scales as $\sigma^{-2}$ and also vanishes. The effective gravity information therefore collapses to zero $F_{\mathrm{eff}}(g,t)\to 0,$ and the correlation coefficient approaches $\rho^2\to 1,$ which is the signature of absolute parameter degeneracy. The free-fall Gaussian benchmark therefore exhibits both the nondegenerate regime, where an internal kinematic timing channel exists, and the singular plane-wave limit, where that channel collapses. 

\section{Discrete metrology: Kasevich--Chu interferometry}
\label{sec:discrete_KC}

We now turn to the canonical three-pulse Kasevich--Chu (KC) atom interferometer and use it to expose the readout dependence of the nuisance-time penalty \cite{KasevichChu1991,KasevichChu1992}. 
In this section, the macroscopic pulse separation is denoted by $T$. The interferometer employs a symmetric $\pi/2$--$\pi$--$\pi/2$ pulse sequence at times $0$, $T$, and $2T$, with an effective wave number $k_0$. The spatial laser phase factor $e^{ik_0\hat z}$ implements the momentum translation $\hat p\mapsto \hat p+\hbar k_0$, and the resulting interference pattern is governed by the familiar phase combination
\begin{equation}
\Delta\Phi(g,T)=-k_0 gT^2-\phi_{\mathrm{ctrl}},
\label{eq:KC_DeltaPhi}
\end{equation}
where $\phi_{\mathrm{ctrl}}$ collects the controllable optical phases. For the pure internal state alone (unit contrast, without motion tracing), the quantum Fisher information for $g$ is $F_{gg}=(\partial_g\Delta\Phi)^2=k_0^2T^4$. The complete pulse algebra and the exact reduction to \eqref{eq:KC_DeltaPhi} are given in Appendix~\ref{app:kc_operator_algebra}. The central point for the present work is that the two-parameter identifiability structure depends strongly on which degrees of freedom are retained in the measurement record.

\subsection{Internal readout: phase-only information and regularized profiling}
\label{subsec:KC_internal_only}

We first consider the standard laboratory setting in which one measures only the internal population and traces out the motional degree of freedom. For a two-outcome fringe with contrast $C\in[0,1]$, the measured likelihood depends on $(g,T)$ only through the single phase variable $\Delta\Phi(g,T)$. The resulting classical Fisher matrix therefore has the outer-product form
\begin{equation}
F^{(\mathrm{int})}_{ij}
=
\mathcal I_{\Phi}\,
(\partial_i\Delta\Phi)(\partial_j\Delta\Phi),
\qquad
i,j\in\{g,T\},
\label{eq:KC_internal_outer_product}
\end{equation}
where $\mathcal I_{\Phi}$ is the phase-information factor determined by the operating point. Because the degree of freedom of motion has been discarded and only the internal outcome statistics are retained, Eq.~\eqref{eq:KC_internal_outer_product} is a classical Fisher matrix rather than a QFI matrix. In mid-fringe, one has $\mathcal I_{\Phi}=C^2$. Equation~\eqref{eq:KC_internal_outer_product} implies
\begin{equation}
F_{gg}^{(\mathrm{int})}F_{TT}^{(\mathrm{int})}
-
\big(F_{gT}^{(\mathrm{int})}\big)^2
=0,
\label{eq:KC_rank1_singularity}
\end{equation}
hence, the internal-only information geometry is exactly rank one. The measured record identifies only the composite combination $gT^2$ and does not separate gravity from the interrogation time. Consequently, the unregularized profiled information vanishes: $F_{\mathrm{eff}}^{(\mathrm{int})}=0. $ A finite gravity variance requires independent timing information to regularize this singular geometry. We model this information through the regularized Schur complement introduced after Eq.~\eqref{eq:Feff_bare}. To emphasize that gravity is estimated after profiling over $T$, we write the effective information here as $F_{g|T}^{(\mathrm{eff,int})}$. Let $I_T^{(\mathrm{prior})}=\frac{1}{(\Delta T_{\mathrm{prior}})^2},$ denote independent timing information supplied by an explicit prior or auxiliary resource. In the intrinsic-clock interpretation emphasized here, one may identify $\Delta T_{\mathrm{prior}}$ with a timing resolution scale constrained by the clock subsystem and, in particular, by quantum speed-limit considerations, as discussed after Eq.~\eqref{eq:Feff_bare} in Sec.~\ref{sec:formalism}.
Using the rank-one identity \eqref{eq:KC_rank1_singularity}, the profiled gravity information becomes
\begin{equation}
F_{g|T}^{(\mathrm{eff,int})}(g,T)
=
F_{gg}^{(\mathrm{int})}
\left(
\frac{I_T^{(\mathrm{prior})}}{F_{TT}^{(\mathrm{int})}+I_T^{(\mathrm{prior})}}
\right).
\label{eq:KC_internal_regularized_general}
\end{equation}
In mid-fringe, the corresponding internal-readout Fisher entries are
\begin{align}
  F_{gg}^{(\mathrm{int})} &= C^2 k_0^2 T^4,
   \\ \nonumber
  F_{TT}^{(\mathrm{int})} &= 4C^2 k_0^2 g^2 T^2,
    \\ \nonumber
  F_{gT}^{(\mathrm{int})} &= 2C^2 k_0^2 g T^3.
    \label{eq:KC_internal_entries_midfringe_gT}
\end{align}
Substituting these expressions into Eq.~\eqref{eq:KC_internal_regularized_general} yields
\begin{equation}
F_{g|T}^{(\mathrm{eff,int})}(g,T)
=
\frac{C^2k_0^2T^4}
{1+4C^2k_0^2g^2T^2(\Delta T_{\mathrm{prior}})^2}.
\label{eq:F_eff_KC_closed}
\end{equation}
Equation~\eqref{eq:F_eff_KC_closed} shows that the internal-only record supports the ideal scaling $T^4$ only when the timing resource is sufficiently sharp. Once the uncertainty of timing dominates, the retained scaling is transferred to an effective $T^2$ behavior.

\subsection{Full-state access and motional timing information}
\label{subsec:KC_full_state}

We now consider the same interferometer at the level of the full final pure state before the motional degree of freedom is discarded. Under ideal closure, the joint output state is factored into a motion component and an internal phase qubit, $\ket{\Psi(g,T)}
=
\ket{\psi_{\mathrm{out}}(T)}
\otimes
\ket{\chi(g,T)},$ where $\ket{\psi_{\mathrm{out}}(T)}$ is independent of $g$ and the internal factor carries the KC phase \eqref{eq:KC_DeltaPhi}. The derivation is given in Appendix~\ref{app:kc_operator_algebra}, in particular subsection~\ref{app:kc_full_state}. This factorization implies that the gravity derivative acts only on the internal phase qubit, while the motional sector contributes an additional timing term. Consequently, the two-parameter information geometry becomes full rank, in contrast to the singular internal-only readout. Because the motion factor is independent of $g$, the gravity derivative acts only on the internal phase qubit. For the gravity parameter, the full-state QFI therefore remains $
F_{gg}^{(\mathrm{diff})}=k_0^2T^4,
$ for one atom under ideal closure, with the obvious factor of $N$ for $N$ independent atoms. 
In the ideal-closure benchmark, the final state is pure and factorized, so the pure-state QFI is additive across the motional and internal factors. The timing sector therefore acquires an additional motional contribution: $F_{TT}^{(\mathrm{diff})}(g,T)
=
F_{TT}^{(\mathrm{int})}(g,T)
+
F_{TT}^{(\mathrm{mot})}(T),$ and the explicit evaluation gives $F_{TT}^{(\mathrm{mot})}(T)
=
{4k_0^2}/{m^2}\operatorname{Var}_{\psi_0}(\hat p).$ The mixed motional terms cancel in the evaluation of the cross element, so the gravity-time coupling is still set by the internal phase derivatives:
$F_{gT}^{(\mathrm{diff})}(g,T)=2k_0^2gT^3.$ Since $F_{TT}^{(\mathrm{mot})}>0$ whenever $\operatorname{Var}_{\psi_0}(\hat p)>0$, the determinant of the complete $2\times2$ QFIM is strictly positive. The complete QFIM is therefore full rank whenever $\operatorname{Var}_{\psi_0}(\hat p)>0$.\\

Introducing the initial longitudinal velocity variance of an atom of mass $m$,
$\sigma_v^2:={\operatorname{Var}_{\psi_0}(\hat p)}/{m^2},$ the unregularized Schur complement for $N$ independent atoms becomes
\begin{equation}
F_{g|T}^{(\mathrm{eff,diff})}(g,T)
=
Nk_0^2T^4
\left[
\frac{\sigma_v^2}{\sigma_v^2+g^2T^2}
\right].
\label{eq:KC_joint_schur_final}
\end{equation}
Equation~\eqref{eq:KC_joint_schur_final} is the discrete interferometric counterpart of the general structural picture developed in Sec.~\ref{sec:kernel_condition_free}. The retained gravity information is stabilized by the momentum variance carried by the initial motional state. In this sense, the motional sector provides a kinematic timing channel that is absent from the phase-only internal record. The physical consequence is immediate. To keep the retention factor in Eq.~\eqref{eq:KC_joint_schur_final} near unity at large $T$, one requires $g^2T^2\ll\sigma_v^2$, equivalently $\sigma_v\gg gT.$ For thermal or weakly collimated ensembles, satisfying the condition $\sigma_v\gg gT$ at macroscopic baselines requires a longitudinal velocity spread large enough to aggravate two standard failure channels of light-pulse interferometry: ballistic expansion relative to the finite Raman beam waist and Doppler detuning relative to the effective Rabi frequency. In that regime, the same momentum variance that preserves geometric identifiability degrades pulse fidelity and fringe contrast. The full-state retention limit should therefore be read as a benchmark bound on the information geometry of the retained state, while practical long-baseline operation relies on auxiliary timing and phase references in addition to the motional channel.

\section{Universal scaling form and platform dictionary}
\label{sec:universal_scaling}

Section~\ref{sec:kernel_condition_free} showed that linearly gravity-coupled sensors with a weak background commutator condition admit a normalized retention law for profiled gravity information. The earlier model sections then provided explicit realizations of that structure. We now collect those realizations into a common scaling description and use it to compare the free-fall benchmark, the prior-regularized phase-only Kasevich--Chu (KC) readout, the full-state KC benchmark, and an idealized closed-system optomechanical benchmark. For the unregularized structural theorem, the effective gravity information takes the form
\begin{align}
  F_{\mathrm{eff}}(g,t) &= F_{gg}(t)\,R(u;\,t),
    \notag\\
  u &:= \frac{g-g_c}{g_*},
    \notag\\
  R(u;\,t) &= 1 - \frac{
    \bigl(\alpha_0(t)+\alpha_1(t)\,u\bigr)^2
  }{1+u^2}.
  \label{eq:universal_retention_recalled}
\end{align}
Here, the normalized coordinate is understood in the domain $g_*>0$, as already specified in Sec.~\ref{sec:kernel_condition_free}. The model dependence enters only through the baseline scale $F_{gg}(t)$, the axis parameters $(g_c,g_*)$, and the normalized numerator coefficients $(\alpha_0,\alpha_1)$. In the KC internal-only problem, the same Lorentzian structure reappears after prior regularization. It is included in the same dictionary because the normalized retention law takes the same Lorentzian form, although its origin lies in a combination of prior regularization and the singular internal‑only readout geometry. The common variable $u$ therefore provides a convenient way to compare benchmark architectures that differ substantially at the dynamical and measurement levels.\\

\begin{figure}[t]
    \centering
    \includegraphics[width=0.88\linewidth]{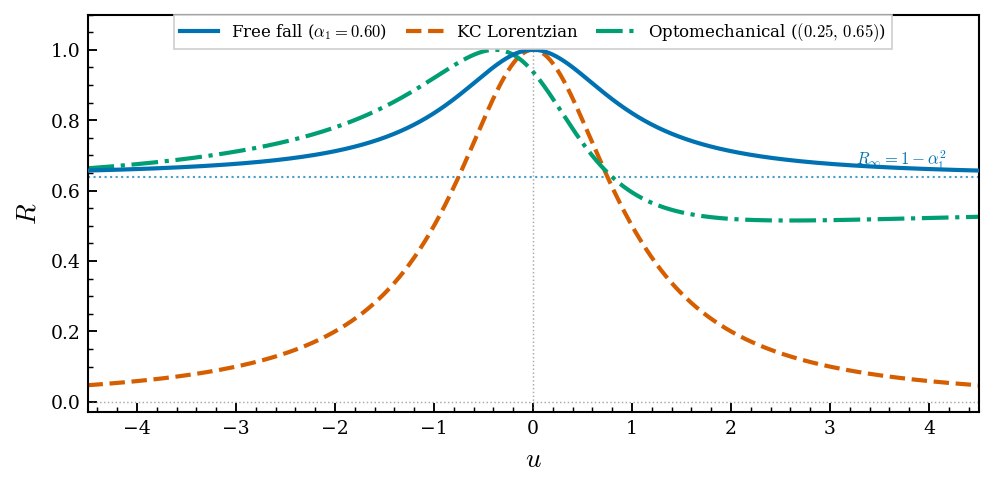}
    \caption{
    Universal normalized nuisance-time retention classes. The horizontal axis is the offset coordinate $u=(g-g_c)/g_*$, and the vertical axis is the retained information fraction $R=F_{\mathrm{eff}}/F_{gg}$. The KC-derived subclasses give the pure Lorentzian profile $R(u)=1/(1+u^2)$.
    The free-fall--type affine-numerator subclass is shown here for $\alpha_0=0$ and a representative admissible value $\alpha_1=0.60$. The optomechanical-type affine-numerator subclass is shown for representative admissible coefficients $\alpha_0=0.25$ and $\alpha_1=0.65$. This figure is structural only: the coefficients are chosen to visualize the distinct normalized kernel shapes implied by Sec.~\ref{sec:kernel_condition_free}, while literature-fixed platform parameters enter instead in Figs.~\ref{fig:opto_periodic_benchmark} and \ref{fig:atom_constrained_regime}.
    }
    \label{fig:universal_retention_classes}
\end{figure}

Figure~\ref{fig:universal_retention_classes} isolates the normalized geometric content of the nuisance-time problem. Its horizontal axis is the offset coordinate $u=(g-g_c)/g_*$, and its vertical axis is the retained information fraction $R=F_{\mathrm{eff}}/F_{gg}$. The panel displays three admissible normalized kernel shapes. The KC-derived subclasses give the pure Lorentzian profile $R(u)=1/(1+u^2)$.
The free-fall benchmark belongs to the $\alpha_0=0$ affine number subclass, for which the retention remains symmetric in $u$ and approaches $1-\alpha_1^2$ at large $|u|$. The optomechanical-type affine-numerator subclass allows $\alpha_0\neq 0$, which shifts the maximum away from $u=0$ and produces an asymmetric profile. Figure~\ref{fig:universal_retention_classes} is therefore structural: it visualizes the distinct normalized kernel classes implied by Sec.~\ref{sec:kernel_condition_free} and does not use literature-fixed platform parameters. Table~\ref{tab:universal_axis_retention} summarizes the dictionary used throughout the paper. The free-fall Gaussian benchmark realizes the general theorem with $\alpha_0=0$ and a nontrivial coefficient $\alpha_1^{\rm ff}(t)$ fixed by the ratio of the $t^2$ kinematic channel to the full gravity information. The prior-regularized internal-only KC record yields the pure Lorentzian form $R(u)=1/(1+u^2)$, with the characteristic scale set by the timing resource carried by $\Delta T_{\mathrm{prior}}$. The full-state KC benchmark yields the same pure Lorentzian retention law, now with the characteristic scale set by the initial velocity spread $\sigma_v$. The optomechanical benchmark is represented in the same affine-numerator form, with platform-dependent coefficients $\alpha_0^{\rm om}(t)$ and $\alpha_1^{\rm om}(t)$. The weak commutator condition and affine cross-term structure for this benchmark are verified in the Appendix~\ref{app:opto_cross_term}. The normalized parameters listed in this row are stated here; their derivation follows from the explicit QFIM evaluation (see, e.g., Ref.~\cite{Wani2026TimeUncertainty}). The optomechanical row in Table~\ref{tab:universal_axis_retention} refers to a closed unitary reference model. In that setting, the mechanical background dynamics remains quadratic, the weak commutator condition is satisfied exactly, and the same kernel structure follows. Practical optomechanical gravimeters operate in an open-system measurement regime, so a full treatment of realistic readout lies beyond the present scope. Its role here is comparative: it shows that the normalized retention structure extends beyond translational free fall and light-pulse interferometry. \\

\begin{figure}[t]
    \centering
    \includegraphics[width=\linewidth]{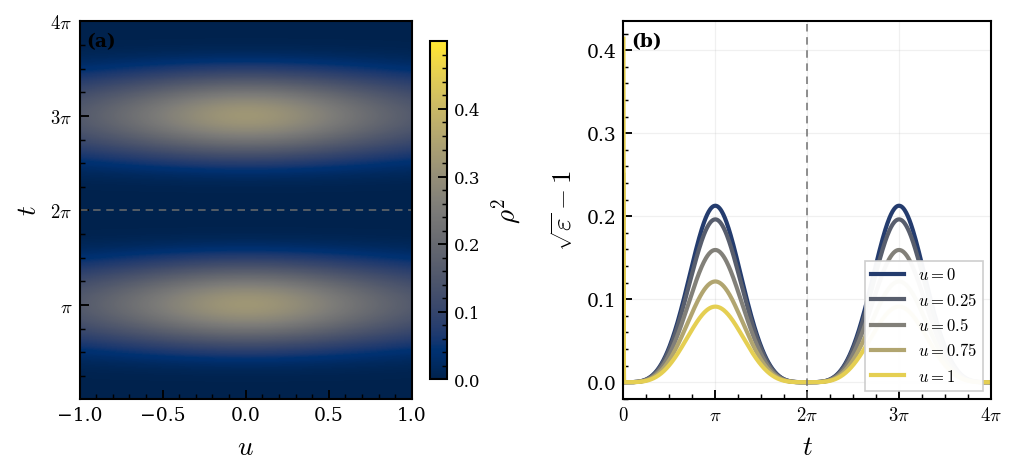}
    \caption{
    Exact closed-unitary optomechanical benchmark in normalized variables. Left: correlation field $\rho^2=F_{gt}^2/(F_{gg}F_{tt})$ plotted as a function of the universal offset coordinate $u=(g-g_c)/g_*$ and the dimensionless time $t=\omega_m t_{\mathrm{phys}}$. Right: relative sensitivity degradation $\sqrt{\varepsilon(u,t)}-1=R(u,t)^{-1/2}-1$ for representative fixed values of $u$. The benchmark family follows the levitated-microobject parameter scale of Ref.~\cite{Qvarfort2018Opto}, while the displayed values $\beta_R=0.10$ and $\beta_I=0$ define an admissible slice chosen for visualization of the exact retention on the plotted domain $u\in[-1,1]$. The dashed vertical lines mark the revival times $t=2\pi n$, where $\sin t=0$ and $1-\cos t=0$ force the gravity--time cross term to vanish.
.
    }
    \label{fig:opto_periodic_benchmark}
\end{figure}

To illustrate how the universal nuisance-time geometry appears in a non-atomic platform, we also show the exact closed-unitary optomechanical benchmark in normalized variables. The benchmark family follows the scale of parameters of the levitated-microobject used in~\cite{Qvarfort2018Opto}, with the same order of magnitude for the mechanical mass, oscillation frequency and number of intracavity photons. Figure~\ref{fig:opto_periodic_benchmark} shows the exact correlation field $\rho^2(u,t)={F_{gt}(u,t)^2}/{F_{gg}(t)\,F_{tt}(u,t)} $ and the corresponding degradation of the relative sensitivity $\sqrt{\varepsilon(u,t)}-1=R(u,t)^{-1/2}-1,$ where $t=\omega_m t_{\mathrm{phys}}$ is the dimensionless mechanical time. The periodic revival structure is exact: at $t=2\pi n$, one has $\sin t=0$ and $1-\cos t=0$, so the gravity--time cross term vanishes and the nuisance-time penalty disappears. The plotted slice uses $\beta_R=0.10$ and $\beta_I=0$ over the displayed interval $u\in[-1,1]$, chosen so that the exact retention remains within the physical range $0\le R\le 1$ throughout the domain shown. The universal axis is therefore best viewed as an organizing variable for the nuisance-time geometry. Once a specific platform is reduced to its $(g_c,g_*)$ scale pair, the gravity information retained is derived from the same normalized function $R(u;t)$. The normalized retention curves discussed here are symmetric-logarithmic-derivative (SLD) information benchmarks. Operational attainability can be more restrictive, though the common retention structure already captures the nuisance-time geometry of the benchmark models analyzed here.

\begin{widetext}

\begin{table}[t]
\centering
\scriptsize
\setlength{\tabcolsep}{4pt}
\renewcommand{\arraystretch}{1.25}
\caption{
Platform dictionary for the normalized nuisance-time retention law. The universal coordinate is
$u=(g-g_c)/g_*$, and the profiled gravity information is written as
$F_{\mathrm{eff}}=F_{gg}R(u;t)$.
The first and fourth rows are direct realizations of the structural theorem of
Sec.~\ref{sec:kernel_condition_free}. The second and third rows are included because they yield the same normalized Lorentzian retention form after, respectively, prior regularization and passage to the full-state KC benchmark. In the free-fall case,
$\alpha_1^{\mathrm{ff}}(t)=\bigl(2m^2\sigma^2 t^2/\hbar^2\bigr)^{1/2}/\sqrt{F_{gg}(t)}$.
For the optomechanical benchmark, $t=\omega_m t_{\mathrm{phys}}$ is dimensionless. In that row,
$\bar k=g_0/\omega_m$, $\mu=\langle \hat a^\dagger\hat a\rangle$,
$\beta=\beta_R+i\beta_I$, $\delta=\Delta_c/\omega_m$, and
$A=\cos\theta\sqrt{m/(2\hbar\omega_m^3)}$, as defined in
Appendix~\ref{app:opto_cross_term}. The appendix establishes the weak commutator condition and affine cross-term structure explicitly, and the detailed reduction to the tabulated normalized optomechanical parameters is quoted here. Under the additional tuning condition
$\delta=-2\bar k\beta_R$, the scale reduces to
$g_*^{\mathrm{om}}=\sqrt{\bar k^2\mu+\beta_I^2}/A$.
}
\label{tab:universal_axis_retention}

\begin{tabular}{lllll}
\toprule
\textbf{Platform}
& \textbf{Axis shift $g_c$}
& \textbf{Scale $g_*$}
& \textbf{Coefficients $(\alpha_0,\alpha_1)$}
& \textbf{Retention $R(u;t)$} \\
\midrule

Free fall (Gaussian benchmark)
&
$0$
&
$\displaystyle \frac{\hbar^2}{2m^2\sigma^3}$
&
$\bigl(0,\alpha_1^{\mathrm{ff}}(t)\bigr)$
&
$\displaystyle 1-\frac{\bigl[\alpha_1^{\mathrm{ff}}(t)\bigr]^2 u^2}{1+u^2}$ \\[6pt]

KC internal-only (prior-regularized)
&
$0$
&
$\displaystyle \frac{1}{2Ck_0T\,\Delta T_{\mathrm{prior}}}$
&
$(0,1)$
&
$\displaystyle \frac{1}{1+u^2}$ \\[6pt]

KC full-state benchmark
&
$0$
&
$\displaystyle \frac{\sigma_v}{T}$
&
$(0,1)$
&
$\displaystyle \frac{1}{1+u^2}$ \\[6pt]

Optomechanical benchmark (closed unitary)
&
$\displaystyle \frac{\bar k\mu-\beta_R}{A}$
&
$\displaystyle \frac{\sqrt{\beta_I^2+\bar k^2\mu+\mu(\delta+2\bar k\beta_R)^2}}{A}$
&
$\bigl(\alpha_0^{\mathrm{om}}(t),\alpha_1^{\mathrm{om}}(t)\bigr)$
&
$\displaystyle 1-\frac{\bigl(\alpha_0^{\mathrm{om}}(t)+\alpha_1^{\mathrm{om}}(t)u\bigr)^2}{1+u^2}$ \\

\bottomrule
\end{tabular}
\end{table}
\end{widetext}

\section{Experimental implications for present atom gravimeters and related atom-interferometric source scales}
\label{sec:experimental_implications}

\begin{figure}[!htbp]
    \centering
    \includegraphics[width=0.88\linewidth]{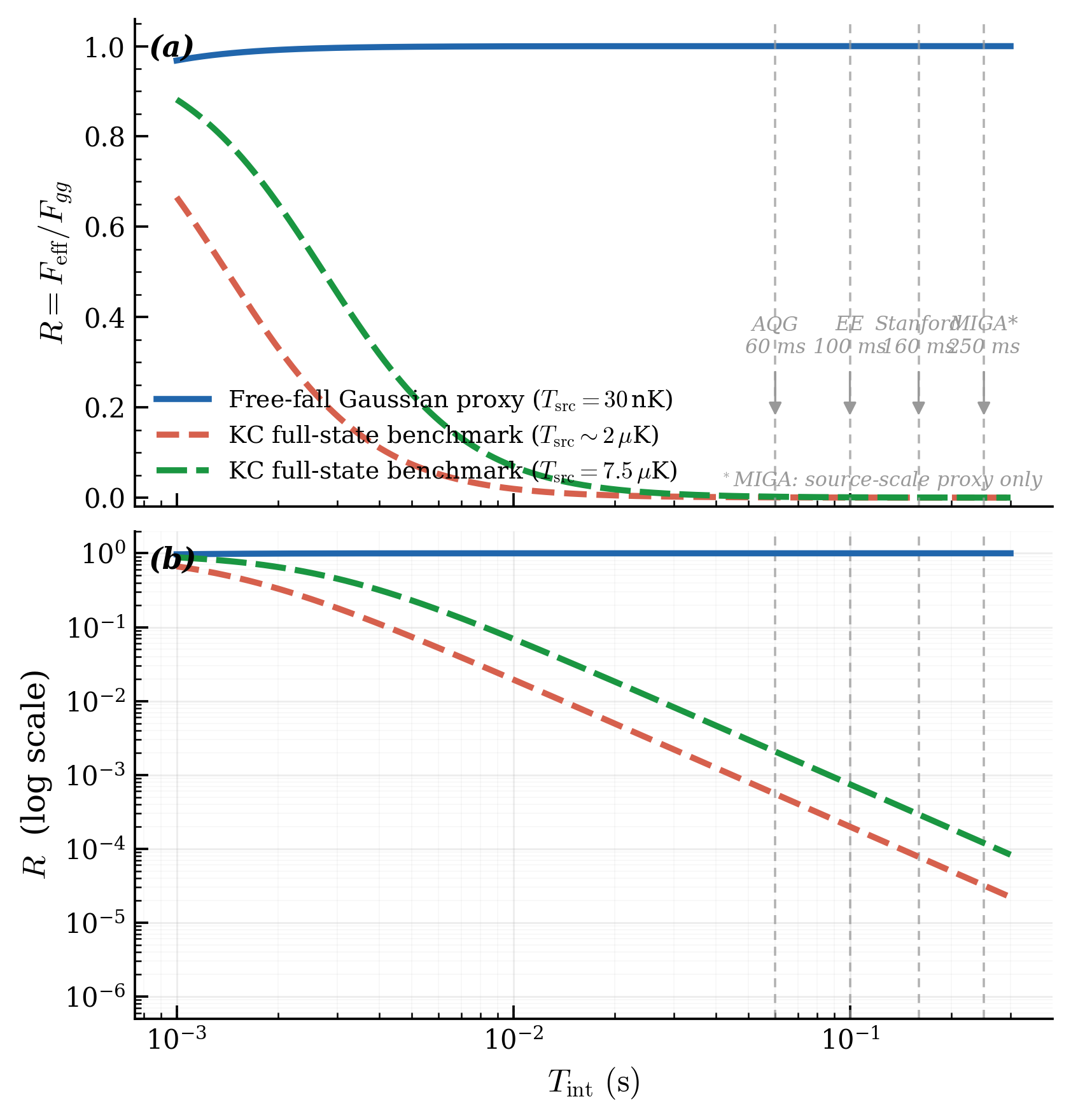}
    \caption{
    Literature-anchored constrained regime for atom-based $^{87}\mathrm{Rb}$ gravimetry. The upper panel shows the retention $R=F_{\mathrm{eff}}/F_{gg}$ on a linear vertical scale, and the lower panel shows the same curves on a logarithmic vertical scale in order to resolve the small-$R$ regime. The orange and green curves use the full-state KC benchmark $R_{\mathrm{KC}}^{(\mathrm{full})}=\sigma_v^2/(\sigma_v^2+g^2T_{\mathrm{int}}^2)$, with one-dimensional rms velocity proxies $\sigma_v=\sqrt{k_B T_{\mathrm{src}}/m_{\mathrm{Rb}}}$ derived from literature-anchored source temperatures representative of the AQG source scale ($T_{\mathrm{src}}\sim2\,\mu\mathrm{K}$) and the Einstein Elevator operating cloud ($T_{\mathrm{src}}=7.5\,\mu\mathrm{K}$) \cite{Menoret2018AQG,Pelluet2025EE}. The MIGA marker at $T_{\mathrm{int}}=250\,\mathrm{ms}$ is included only as a source-scale comparison point based on the $\sim2\,\mu\mathrm{K}$ source temperature; the actual velocity-selected interferometer-axis input in that platform is much narrower \cite{Beaufils2022MIGA}. The blue curve is a free-fall Gaussian proxy obtained by mapping the same momentum-width scale to a minimum-uncertainty Gaussian benchmark using $\sigma=\frac{\hbar}{\sqrt{2}\,m_{\mathrm{Rb}}\,\sigma_v^{(\mathrm{rms})}}$; the comparison value $T_{\mathrm{src}}=30\,\mathrm{nK}$ is taken from the ultracold-source capability reported on the Einstein Elevator apparatus \cite{Pelluet2025EE}. Vertical markers indicate representative interrogation times from AQG ($60\,\mathrm{ms}$), the Einstein Elevator sensitivity point ($100\,\mathrm{ms}$), the Stanford fountain benchmark ($160\,\mathrm{ms}$), and the MIGA design scale ($250\,\mathrm{ms}$) \cite{Menoret2018AQG,Pelluet2025EE,Peters1999Nature,Beaufils2022MIGA}. Here $T_{\mathrm{int}}$ denotes interrogation time and $T_{\mathrm{src}}$ denotes source temperature. The plotted values are benchmark mappings of literature source scales onto the retention laws and are not direct reconstructions of the measured operating sensitivities of the marked instruments.
    }
    \label{fig:atom_constrained_regime}
\end{figure}

We now translate the full-state Kasevich--Chu retention law into the parameter regime of present light-pulse atom gravimeters. The resulting mapping isolates the fraction of retained gravity information generated by unperturbed longitudinal motion spread alone, without auxiliary timing information in the retained record. To place the detailed present-baseline estimates in a broader atom-interferometric source-scale context, Fig.~\ref{fig:atom_constrained_regime} maps the benchmark retention laws using literature-anchored $^{87}\mathrm{Rb}$ parameters. For the transportable absolute quantum gravimeter (AQG), we use the reported pulse separation $T=60\,\mathrm{ms}$ together with cooling to about $2\,\mu\mathrm{K}$ \cite{Menoret2018AQG}. For the Einstein Elevator platform, we use the measured operating-cloud temperature $T_{\mathrm{src}}=7.5\,\mu\mathrm{K}$, the demonstrated sensitivity point at total interferometer time $2T=200\,\mathrm{ms}$ (pulse separation $T=100\,\mathrm{ms}$), and the reported extension of coherent operation to $2T=260\,\mathrm{ms}$ ($T=130\,\mathrm{ms}$) \cite{Pelluet2025EE}. For the MIGA cold-atom source, we use the source-scale temperature $\approx 2\,\mu\mathrm{K}$ and the design pulse separation $T=250\,\mathrm{ms}$ of the future horizontal Bragg interferometer \cite{Beaufils2022MIGA}.  In that case, however, the actual velocity-selected distribution relevant to the interferometer axis is much narrower than the source-scale value, so the MIGA point in Fig.~\ref{fig:atom_constrained_regime} should only be read as a source-scale comparison marker rather than as an operating interferometer-input estimate \cite{Beaufils2022MIGA}. As a historical long-$T$ reference, we also mark the Doppler-sensitive Stanford fountain benchmark at $T_{\mathrm{int}}=160\,\mathrm{ms}$ \cite{Peters1999Nature}. We place these literature values on the benchmark retention curves by converting each reported source temperature into a one-dimensional rms velocity proxy, $\sigma_v^{(\mathrm{rms})}
=
\sqrt{{k_B T_{\mathrm{src}}}/{m_{\mathrm{Rb}}}}$.
Here $\sigma_v^{(\mathrm{rms})}$ is used only as a one-dimensional source-scale proxy for the longitudinal velocity width. It is not intended to reconstruct the exact interferometer-input distribution for platforms that apply additional velocity selection or other state-preparation steps along the measurement axis. We then insert this width into the full-state KC benchmark expression
\begin{equation}
R_{\mathrm{KC}}^{(\mathrm{full})}(g,T_{\mathrm{int}})
=
\frac{\sigma_v^2}{\sigma_v^2+g^2T_{\mathrm{int}}^2}.
\label{eq:KC_retention_experimental}
\end{equation}
For the free-fall Gaussian benchmark, we use the same momentum-width scale and map it to a minimum-uncertainty Gaussian proxy via $\sigma
=
{\hbar}/{\sqrt{2}\,m_{\mathrm{Rb}}\sigma_v^{(\mathrm{rms})}}.$ This relation follows from the minimum-uncertainty condition $\operatorname{Var}_{\psi_0}(\hat p)=\hbar^2/(2\sigma^2)$ and $\sigma_v^{(\mathrm{rms})}=\sqrt{\operatorname{Var}_{\psi_0}(\hat p)}/m$. For a fixed $\sigma_v$, this choice yields the smallest possible position variance and therefore represents an upper bound on the retention achievable by a Gaussian state with that momentum spread. The comparison value $T_{\mathrm{src}}=30\,\mathrm{nK}$ is taken from the ultracold-source capability reported on the Einstein Elevator apparatus and is not the operating temperature of the microgravity interferometer dataset itself \cite{Pelluet2025EE}. Figure~\ref{fig:atom_constrained_regime} therefore provides a literature-anchored benchmark mapping from source scales and interrogation times to retention laws, rather than a reconstruction of the full sensitivity at the operating point or the longitudinal input state of each interferometer platform.\\
Figure~\ref{fig:atom_constrained_regime} shows that source-scale microkelvin widths place the full-state KC benchmark, when no additional timing information is included, deep in the low-retention regime once the interrogation time approaches the $0.1-0.25\,\mathrm{s}$ scale, while the free-fall Gaussian proxy remains close to unity and approaches complete retention over the same interval. We now make this scale concrete for representative present Raman gravimeters. For concrete order-of-magnitude estimates, two representative present Raman-gravimeter baselines are particularly useful. The transportable AQG instrument operates with $^{87}$Rb atoms, Raman interrogation time $T=60\,\mathrm{ms}$, source cooling to approximately $2\,\mu\mathrm{K}$, and internal-state fluorescence detection \cite{Menoret2018AQG}. The mobile fountain gravimeter GAIN likewise uses $^{87}$Rb Raman interferometry, with pulse-separation time $T=260\,\mathrm{ms}$, narrow vertical velocity selection, and state-sensitive fluorescence detection \cite{GAINOverview,Freier2016}. Therefore, these instruments realize high-performance internal-readout gravimetry with explicit experimental timing, phase-control, and auxiliary reference resources. They therefore lie outside the bare full-state KC benchmark analyzed in Sec.~\ref{subsec:KC_full_state}, in which the retained probe state alone is asked to distinguish the gravity and interrogation-time directions, so the benchmark retentions estimated below should not be read as reconstructions of their measured operating sensitivities. For order-of-magnitude estimates, we take $T_{\mathrm{src}}\simeq 2\,\mu\mathrm{K}$ as a representative microkelvin source scale for the present $^{87}$Rb Raman gravimeters. In the case of GAIN, which applies narrow vertical velocity selection before the interferometer, this should be viewed as a generous upper-bound proxy for the longitudinal input width. The corresponding thermal velocity scale is
\begin{equation}
\sigma_v^{(\mathrm{th})}
=
\sqrt{\frac{k_B T_{\mathrm{at}}}{m_{\mathrm{Rb}}}}.
\end{equation}
Using $m(^{87}\mathrm{Rb})\simeq 1.443\times 10^{-25}\,\mathrm{kg}$ gives
\begin{equation}
\sigma_v^{(\mathrm{th})}
\simeq 1.38\times 10^{-2}\,\mathrm{m\,s^{-1}}.
\end{equation} For AQG, with $T=60\,\mathrm{ms}$ and $gT\simeq 5.89\times 10^{-1}\,\mathrm{m\,s^{-1}}$, this gives
\begin{equation}
R_{\mathrm{AQG}}^{(\mathrm{th})}
=
\frac{\bigl(\sigma_v^{(\mathrm{th})}\bigr)^2}{\bigl(\sigma_v^{(\mathrm{th})}\bigr)^2+g^2T^2}
\simeq 5.5\times 10^{-4}.
\label{eq:R_AQG_estimate}
\end{equation}
For GAIN, with $T=260\,\mathrm{ms}$ and $gT\simeq 2.55\,\mathrm{m\,s^{-1}}$, we find $R_{\mathrm{GAIN}}^{(\mathrm{th})}
 \simeq 2.9\times 10^{-5}.$ For GAIN, these values should be interpreted as generous upper-bound retention estimates predicted by the bare full-state KC benchmark at that baseline, because narrow vertical velocity selection makes the actual longitudinal interferometer-input width smaller than the source-scale proxy used here. These values show that, at the interrogation times already realized in present $^{87}$Rb Raman gravimeters, the motional spread associated with microkelvin atom sources is much smaller than the scale $gT$ required for strong retention within the bare full-state KC benchmark, where the motional sector provides the only contribution to separating the time direction from the gravity direction. The corresponding nuisance-time penalty is therefore already large in that benchmark regime, even though it is not directly visible in standard instrument operation, because present gravimeters do not rely on the motional sector alone to supply the information that separates interrogation time from gravity in the retained record. The same conclusion appears from the inverse requirement. Imposing a target retention level $R_0\in(0,1)$ gives the following.
\begin{equation}
\sigma_v
\ge
gT\sqrt{\frac{R_0}{1-R_0}}
=: \alpha(R_0)\,gT.
\label{eq:retention_requirement_experiment}
\end{equation}
For AQG, half-retention ($R_0=1/2$) requires $\sigma_v\simeq 5.89\times 10^{-1}\,\mathrm{m\,s^{-1}}$, while retention $90\%$ requires $\sigma_v\simeq 1.77\,\mathrm{m\,s^{-1}}$. For GAIN, the corresponding values are $\sigma_v\simeq 2.55\,\mathrm{m\,s^{-1}}$ and $\sigma_v\simeq 7.65\,\mathrm{m\,s^{-1}}$. These are higher than the $2\,\mu\mathrm{K}$ thermal scale by factors of about $43$ and $128$ for AQG and about $184$ and $553$ for GAIN. For transform-limited pure states satisfying $\Delta z_0 \Delta p=\hbar/2$, Eq.~\eqref{eq:retention_requirement_experiment} implies $\Delta z_0
\le
{\hbar}/{2m\,\alpha(R_0)\,gT}.$ This gives longitudinal localization bounds of approximately $621\,\mathrm{pm}$ and $207\,\mathrm{pm}$ on the AQG baseline for $R_0=1/2$ and $R_0=0.9$, and about $143\,\mathrm{pm}$ and $47.7\,\mathrm{pm}$ on the GAIN baseline. These are Heisenberg-saturating pure-state benchmark bounds, not realistic cloud sizes for present thermal Raman gravimeters. The corresponding pure-state preparation requirement is therefore pushed into a sub-nanometer localization regime, together with the associated Doppler and expansion costs quantified in the present section.\\

This comparison also clarifies which experimental upgrades can and cannot mitigate the nuisance-time penalty within the bare full-state KC benchmark. Increasing the interrogation time improves the nominal single-parameter phase sensitivity, but at fixed $\sigma_v$ it decreases the retained gravity information because $R_{\mathrm{KC}}^{(\mathrm{full})}$ falls once $gT\gg \sigma_v$. Large-momentum-transfer beam splitters and quantum-enhanced atom-number scaling can improve the ideal prefactor of the gravimetric sensitivity \cite{FloquetLMT2024,Fuderer2023}, yet within the full-state Kasevich--Chu benchmark, they do not alter the retention factor itself, which depends only on the ratio $\sigma_v/(gT)$. In this sense, larger $k_{\mathrm{eff}}$ or larger $N$ amplify the nominal signal channel, while the gravity--time identifiability bottleneck remains controlled by the motional contribution to the time-information block.  In the present benchmark framework, the strong performance of Raman gravimeters is consistent with architectures that use internal-state readout, laser phase control, and external timing or phase references. These resources supply the information needed to separate interrogation time from gravity in the retained record. The full-state KC benchmark without auxiliary timing information would instead require longitudinal motional spreads that are far beyond the microkelvin source scales of current instruments. Our framework thus identifies, in explicit order-of-magnitude terms, the state-preparation and time-distinguishability burden hidden behind the ideal $T^4$ scaling law once gravity and interrogation time are treated on equal footing.

\section{Conclusion}
\label{sec:conclusion}

Within the pure-state unitary linearly gravity-coupled setting studied here, inherently uncertain interrogation time turns quantum gravimetry into a two-parameter estimation problem. In this setting, the nuisance-time penalty acquires a rigid geometric structure: the timing block of the QFIM is quadratic in the gravitational parameter, and, under the weak commutator condition, the gravity--time cross term is affine. The gravity information that survives profiling over time therefore follows a normalized retention law with a Lorentzian kernel.  For the freely falling Gaussian probe, profiling over time leaves an exactly retained $t^4$ contribution while suppressing only the $t^2$ channel. For the Kasevich--Chu interferometer with internal-only readout, gravity and interrogation time collapse into a single phase variable, so independent timing information is required to regularize the two-parameter geometry. In the full-state KC benchmark, the retention of motional information restores a full-rank estimation problem, with the retained gravity information controlled by the initial velocity variance. The idealized optomechanical benchmark shows that the same retention structure extends to quadratic background dynamics beyond translational free fall. The main implication is that nominal gravity sensitivity and gravity--time identifiability must be treated as distinct resources in this setting. The nearly ideal $T^4$ scaling survives nuisance-time profiling only when the sensing architecture preserves an independent timing channel or when the timing information is supplied explicitly. In the differential KC setting, that requirement becomes increasingly costly at long baselines because the needed momentum spread competes with localization, free expansion, Doppler broadening, and interferometric visibility. A natural next step is to extend the same nuisance-time analysis beyond the closed unitary benchmarks considered here to realistic open-system readout and explicit clock-assisted architectures. Within the scope of the models analyzed in this paper, the familiar single-parameter scaling laws of quantum gravity therefore overstate what remains once interrogation time is estimated rather than assumed.
\section*{Acknowledgement}
This publication was partially supported by the Qatar Research, Development and Innovation (QRDI) Council under the Academic Research Grant ARG01-0603-230468. The findings and views expressed herein are solely the responsibility of the authors.

\bibliographystyle{apsrev4-2}
\bibliography{main}
\clearpage
\onecolumngrid

\appendix

\section{Affine gravity--time cross term under the weak-commutator condition}
\label{app:dyson_proof}

This appendix proves the structural statement used in Sec.~\ref{sec:kernel_condition_free}. Assume the $g$-independent dynamics satisfy the weak-commutator condition
\begin{equation}
[\hat z_0(u),\hat z_0(v)]
=
i\hbar\,\Delta(u,v)\,\mathbb I,
\qquad
u,v\in[0,t],
\label{eq:app_weak_commutator_criterion}
\end{equation}
with $\Delta(u,v)\in\mathbb R$. Then the local gravity generator differs from its $g=0$ value only by a scalar identity shift, and consequently the QFIM cross element $F_{gt}(g,t)$ is affine in $g$.

We consider the linear gravity-coupled Hamiltonian
\begin{equation}
\hat H(g)=\hat H_0+m g\,\hat z.
\end{equation}
Let
\begin{equation}
\hat z_0(s):=
e^{+\frac{i}{\hbar}\hat H_0 s}\,\hat z\,e^{-\frac{i}{\hbar}\hat H_0 s}
\end{equation}
denote the free-evolved position operator, and define the interaction-picture propagator by
\begin{equation}
\hat U_I(g,s):=
e^{+\frac{i}{\hbar}\hat H_0 s}\,\hat U(g,s),
\qquad
\hat U(g,s)=e^{-i\hat H(g)s/\hbar},
\end{equation}
so that a direct differentiation gives
\begin{equation}
\partial_s \hat U_I(g,s)
=
\hat Z(s)\,\hat U_I(g,s),
\qquad
\hat Z(s):=-\frac{i g m}{\hbar}\,\hat z_0(s),
\end{equation}
hence
\begin{equation}
\hat U_I(g,s)
=
\mathcal T\exp\!\left(\int_0^s \hat Z(\tau)\,d\tau\right).
\end{equation}

According to Eq.~\eqref{eq:app_weak_commutator_criterion}, the interaction-picture generator has a c-number commutator:
\begin{equation}
[\hat Z(\tau_1),\hat Z(\tau_2)]
=
-\frac{g^2m^2}{\hbar^2}
[\hat z_0(\tau_1),\hat z_0(\tau_2)]
=
-\frac{i g^2m^2}{\hbar}\,\Delta(\tau_1,\tau_2)\,\mathbb I,
\end{equation}
for all $\tau_1,\tau_2\in[0,s]$. Therefore all higher nested commutators vanish, and the Magnus expansion truncates after the second term:
\begin{equation}
\hat U_I(g,s)=e^{\Omega_1(s)+\Omega_2(s)},
\end{equation}
with
\begin{equation}
\Omega_1(s)=\int_0^s \hat Z(\tau_1)\,d\tau_1,
\qquad
\Omega_2(s)=\frac12\int_0^s\!\!\int_0^{\tau_1}
[\hat Z(\tau_1),\hat Z(\tau_2)]\,d\tau_2\,d\tau_1.
\end{equation}
Since $\Omega_2(s)$ is itself a $c$-number, it commutes with $\Omega_1(s)$, so
\begin{equation}
\hat U_I(g,s)=e^{\Omega_1(s)}e^{\Omega_2(s)}.
\end{equation}
The second factor is a global scalar phase.

The full Heisenberg position operator under $\hat H(g)$ may be written as
\begin{equation}
\hat z_H(s;g)
=
\hat U_I^\dagger(g,s)\,\hat z_0(s)\,\hat U_I(g,s).
\end{equation}
Because the scalar factor $e^{\Omega_2(s)}$ cancels in conjugation, only $e^{\Omega_1(s)}$ contributes:
\begin{equation}
\hat z_H(s;g)
=
e^{-\Omega_1(s)}\,\hat z_0(s)\,e^{+\Omega_1(s)}.
\end{equation}
Applying the Hadamard expansion,
\begin{equation}
e^{-X}Ye^{X}
=
Y+[Y,X]+\frac{1}{2!}[[Y,X],X]+\cdots,
\end{equation}
with $X=\Omega_1(s)$ and $Y=\hat z_0(s)$, we obtain
\begin{align}
\hat z_H(s;g)
&=
\hat z_0(s)
+
\int_0^s [\hat z_0(s),\hat Z(\tau_1)]\,d\tau_1
+
\frac{1}{2!}\int_0^s\!\!\int_0^s
[[\hat z_0(s),\hat Z(\tau_1)],\hat Z(\tau_2)]\,d\tau_2\,d\tau_1
+\cdots.
\end{align}
The first commutator is
\begin{equation}
[\hat z_0(s),\hat Z(\tau_1)]
=
-\frac{i g m}{\hbar}
[\hat z_0(s),\hat z_0(\tau_1)]
=
- g m\,\Delta(\tau_1,s)\,\mathbb I,
\end{equation}
which is already proportional to the identity. Consequently every higher nested commutator vanishes identically. The series truncates at first order, yielding
\begin{equation}
\hat z_H(s;g)
=
\hat z_0(s)+g\,f(s)\,\mathbb I,
\qquad
f(s):=-m\int_0^s \Delta(\tau_1,s)\,d\tau_1\in\mathbb R.
\end{equation}

The local gravity generator is
\begin{equation}
\hat{\mathcal H}_g(g,t)
=
\frac{m}{\hbar}\int_0^t \hat z_H(s;g)\,ds.
\end{equation}
Substituting the affine form of $\hat z_H$ gives
\begin{equation}
\hat{\mathcal H}_g(g,t)
=
\hat{\mathcal H}_g(0,t)+g\,\kappa(t)\,\mathbb I,
\qquad
\kappa(t)=\frac{m}{\hbar}\int_0^t f(s)\,ds.
\end{equation}

Recall from Sec.~\ref{sec:kernel_condition_free} that
\begin{equation}
\hat{\mathcal H}_t(g,t)=\hat A+g\hat B,
\qquad
\hat A=\frac{\hat H_0}{\hbar},
\qquad
\hat B=\frac{m}{\hbar}\hat z.
\end{equation}
Since $\operatorname{Cov}^{\mathrm{sym}}_{\psi_0}(\mathbb I,\hat A)=\operatorname{Cov}^{\mathrm{sym}}_{\psi_0}(\mathbb I,\hat B)=0$, 
we obtain
\begin{equation}
\begin{aligned}
F_{gt}(g,t)
&=
4\,\operatorname{Cov}^{\mathrm{sym}}_{\psi_0}\!\big(\hat{\mathcal H}_g(g,t),\hat{\mathcal H}_t(g,t)\big)
\\
&=
4\,\operatorname{Cov}^{\mathrm{sym}}_{\psi_0}\!\big(\hat{\mathcal H}_g(0,t)+g\kappa(t)\mathbb I,\hat A+g\hat B\big)
\\
&=
4\,\operatorname{Cov}^{\mathrm{sym}}_{\psi_0}\!\big(\hat{\mathcal H}_g(0,t),\hat A\big)
+
4g\,\operatorname{Cov}^{\mathrm{sym}}_{\psi_0}\!\big(\hat{\mathcal H}_g(0,t),\hat B\big)
\\
&=:d_0(t)+d_1(t)\,g.
\end{aligned}
\label{eq:app_Fgt_affine_final}
\end{equation}
This is the affine form used in the main text.

\section{Exact BCH factorization and QFIM evaluation for Gaussian free fall}
\label{app:bch_algebra}

This appendix derives the exact factorization and QFIM entries used in Sec.~\ref{sec:continuous_free_fall} for the free-fall Hamiltonian
\begin{equation}
\hat H(g)=\frac{\hat p^2}{2m}+mg\,\hat z.
\end{equation}

\subsection{Exact BCH factorization}
\label{app:bch_factorization}

Write
\begin{equation}
\hat U(g,t)=e^{\hat X+\hat Y},
\qquad
\hat X:=-\frac{i t}{\hbar}\frac{\hat p^2}{2m},
\qquad
\hat Y:=-\frac{i t}{\hbar}mg\,\hat z.
\end{equation}
Using $[\hat z,\hat p]=i\hbar$, we first compute $[\hat p^2,\hat z]=\hat p[\hat p,\hat z]+[\hat p,\hat z]\hat p=-2i\hbar\,\hat p$.
Hence
\begin{align}
[\hat X,\hat Y]
&=
\left[-\frac{i t}{\hbar}\frac{\hat p^2}{2m},-\frac{i t}{\hbar}mg\,\hat z\right]
=
\frac{i g t^2}{\hbar}\,\hat p,
\\
[\hat Y,[\hat X,\hat Y]]
&=
\left[-\frac{i t}{\hbar}mg\,\hat z,\frac{i g t^2}{\hbar}\hat p\right]
=
\frac{i m g^2 t^3}{\hbar}.
\end{align}
Since $[\hat Y,[\hat X,\hat Y]]$ is proportional to the identity, all higher nested commutators vanish. The BCH series therefore truncates, and one obtains
\begin{equation}
\hat U(g,t)
=
e^{-\frac{i t}{\hbar}\frac{\hat p^2}{2m}}\,
e^{-\frac{i t}{\hbar}mg\,\hat z}\,
e^{-\frac{i g t^2}{2\hbar}\hat p}\,
e^{+\frac{i m g^2 t^3}{3\hbar}}.
\label{eq:app_Ufactor1}
\end{equation}

We now combine the linear $\hat z$ and $\hat p$ exponentials. Define $\hat A_1:=-(it/\hbar)mg\,\hat z$ and $\hat A_2:=-ig t^2\hat p/(2\hbar)$.
Their commutator is $[\hat A_1,\hat A_2]=-i m g^2 t^3/(2\hbar)$, 
again a scalar. Therefore
\begin{equation}
e^{\hat A_1}e^{\hat A_2}
=
e^{\hat A_1+\hat A_2}\,e^{\frac12[\hat A_1,\hat A_2]}
=
\exp\!\left[
-\frac{i g t}{\hbar}
\left(
m\hat z+\frac{t}{2}\hat p
\right)
\right]
\exp\!\left(-\frac{i m g^2 t^3}{4\hbar}\right).
\end{equation}
Substituting into Eq.~\eqref{eq:app_Ufactor1}, the scalar phases combine as $1/3-1/4=1/12$,
so
\begin{equation}
\hat U(g,t)
=
\exp\!\left[-\frac{i t}{\hbar}\frac{\hat p^2}{2m}\right]
\exp\!\left[-ig\,\hat G_0(t)\right]
\exp\!\left[+\frac{i m g^2 t^3}{12\hbar}\right],
\label{eq:app_Ufinal}
\end{equation}
with
\begin{equation}
\hat G_0(t)=\frac{t}{\hbar}\left(m\hat z+\frac{t}{2}\hat p\right).
\label{eq:app_G0_def}
\end{equation}

\subsection{QFIM entries for the centered Gaussian}
\label{app:qfim_gaussian}

Take the centered minimum-uncertainty Gaussian state with
\begin{equation}
\operatorname{Var}_{\psi_0}(\hat z)=\frac{\sigma^2}{2},
\qquad
\operatorname{Var}_{\psi_0}(\hat p)=\frac{\hbar^2}{2\sigma^2},
\qquad
\operatorname{Cov}^{\mathrm{sym}}_{\psi_0}(\hat z,\hat p)=0.
\end{equation}
The QFIM entries are $F_{ij}=4\,\operatorname{Cov}^{\mathrm{sym}}_{\psi_0}\!\big(\hat{\mathcal H}_i,\hat{\mathcal H}_j\big)$.

From Eq.~\eqref{eq:app_Ufinal}, the gravity generator is
\begin{equation}
\hat{\mathcal H}_g=\hat G_0(t)=\frac{t}{\hbar}\left(m\hat z+\frac{t}{2}\hat p\right),
\end{equation}
so
\begin{equation}
\begin{aligned}
F_{gg}
&=
4\,\operatorname{Var}_{\psi_0}\!\left(
\frac{t}{\hbar}\left(m\hat z+\frac{t}{2}\hat p\right)
\right)
\\
&=
\frac{4t^2}{\hbar^2}
\left(
m^2\operatorname{Var}_{\psi_0}(\hat z)
+\frac{t^2}{4}\operatorname{Var}_{\psi_0}(\hat p)
+m t\,\operatorname{Cov}^{\mathrm{sym}}_{\psi_0}(\hat z,\hat p)
\right)
\\
&=
\frac{t^4}{2\sigma^2}
+
\frac{2m^2\sigma^2 t^2}{\hbar^2}.
\end{aligned}
\label{eq:app_Fgg_final}
\end{equation}

For the time generator,
\begin{equation}
\hat{\mathcal H}_t=\frac{\hat H(g)}{\hbar}
=
\frac{1}{\hbar}\left(\hat T+\hat V\right),
\qquad
\hat T:=\frac{\hat p^2}{2m},
\qquad
\hat V:=mg\,\hat z.
\end{equation}
Hence
\begin{equation}
F_{tt}
=
\frac{4}{\hbar^2}\operatorname{Var}_{\psi_0}(\hat T+\hat V).
\label{eq:app_Ftt_start}
\end{equation}
For the centered Gaussian, the mixed term vanishes: $\operatorname{Cov}^{\mathrm{sym}}_{\psi_0}(\hat T,\hat V)=0$,
so $\operatorname{Var}_{\psi_0}(\hat T+\hat V)=\operatorname{Var}_{\psi_0}(\hat T)+\operatorname{Var}_{\psi_0}(\hat V)$. 
For a zero-mean Gaussian in momentum, $\langle \hat p^4\rangle_{\psi_0}=3\langle \hat p^2\rangle_{\psi_0}^2$ and $\operatorname{Var}_{\psi_0}(\hat p^2)=2\,\operatorname{Var}_{\psi_0}(\hat p)^2$.
Therefore
\begin{equation}
\begin{aligned}
\operatorname{Var}_{\psi_0}(\hat T)
&=
\frac{1}{4m^2}\operatorname{Var}_{\psi_0}(\hat p^2)
=
\frac{\hbar^4}{8m^2\sigma^4},
\\
\operatorname{Var}_{\psi_0}(\hat V)
&=
m^2g^2\operatorname{Var}_{\psi_0}(\hat z)
=
\frac{m^2g^2\sigma^2}{2}.
\end{aligned}
\label{eq:app_varT_varV}
\end{equation}
Substituting into Eq.~\eqref{eq:app_Ftt_start},
\begin{equation}
F_{tt}
=
\frac{\hbar^2}{2m^2\sigma^4}
+
\frac{2m^2g^2\sigma^2}{\hbar^2}.
\label{eq:app_Ftt_final}
\end{equation}

Finally, $F_{gt}=4\,\operatorname{Cov}^{\mathrm{sym}}_{\psi_0}\!\left(\hat{\mathcal H}_g,\hat H(g)/\hbar\right)$.
The mixed covariance with $\hat T$ vanishes by symmetry, so only the potential term contributes:
$F_{gt}=(4mg/\hbar)\,\operatorname{Cov}^{\mathrm{sym}}_{\psi_0}(\hat G_0(t),\hat z)$. 
Using Eq.~\eqref{eq:app_G0_def},
\begin{equation}
\begin{aligned}
\operatorname{Cov}^{\mathrm{sym}}_{\psi_0}(\hat G_0(t),\hat z)
&=
\frac{t}{\hbar}
\left(
m\operatorname{Var}_{\psi_0}(\hat z)
+\frac{t}{2}\operatorname{Cov}^{\mathrm{sym}}_{\psi_0}(\hat p,\hat z)
\right)
\\
&=
\frac{t}{\hbar}\,m\,\frac{\sigma^2}{2}.
\end{aligned}
\end{equation}
Therefore
\begin{equation}
F_{gt}
=
\frac{2m^2g\sigma^2 t}{\hbar^2}.
\label{eq:app_Fgt_final}
\end{equation}

Collecting Eqs.~\eqref{eq:app_Fgg_final}, \eqref{eq:app_Ftt_final}, and \eqref{eq:app_Fgt_final} reproduces the QFIM in Eq.~\eqref{eq:QFIM_Gaussian}. This benchmark also realizes the universal kernel with $g_c=0$, $\alpha_0=0$, and $g_*=\hbar^2/(2m^2\sigma^3)$.

\section{Optomechanical benchmark: weak-condition verification and affine cross term}
\label{app:opto_cross_term}

This appendix records the optomechanical benchmark ingredients used in the structural classification of Sec.~\ref{sec:universal_scaling}. The material established here is limited to two points: verification of the weak commutator condition and the affine dependence of the gravity--time cross term on $g$.

We consider the closed-system benchmark Hamiltonian in the rotating frame,
\begin{equation}
\hat H
=
-\hbar\Delta_c\,\hat a^\dagger \hat a
+
\hbar\omega_m\,\hat b^\dagger \hat b
-
\hbar g_0(\hat b^\dagger+\hat b)\hat a^\dagger \hat a
+
mg\,\hat z.
\label{eq:app_opto_H}
\end{equation}
Here $\Delta_c$ is the cavity detuning, $\omega_m$ is the mechanical frequency, $g_0$ is the single-photon optomechanical coupling, $\mu=\langle \hat a^\dagger\hat a\rangle$ is the mean number of photons in the cavity, $\beta=\beta_R+i\beta_I$ is the initial coherent mechanical amplitude, and
\begin{equation}
\bar k:=\frac{g_0}{\omega_m},
\qquad
\delta:=\frac{\Delta_c}{\omega_m},
\qquad
\zeta(t):=t-\sin t,
\qquad
A:=\cos\theta\sqrt{\frac{m}{2\hbar\omega_m^3}}.
\end{equation}

The radiation-pressure interaction commutes with the number of cavity photons, $[\hat a^\dagger\hat a,\hat H]=0$,
so within this reference the optical occupation is conserved. The mechanical background dynamics is quadratic, generated by $\hbar\omega_m \hat b^\dagger \hat b$. Consequently, the free-evolved position operator $\hat z_0(s)$ remains a linear combination of $\hat z$ and $\hat p$, and hence
$[\hat z_0(u),\hat z_0(v)]\propto \sin[\omega_m(v-u)]\,\mathbb I$
is a $c$-number for all $u,v$. The weak commutator condition of Appendix~\ref{app:dyson_proof} is therefore satisfied exactly.

For the coherent-state benchmark used in the main text, the closed-form evaluation of the QFIM yields the cross element
\begin{equation}
F_{gt}(g,t)
=
A\Big[
-16\,\bar k^2\mu\,\zeta(t)\,(1-\cos t)\,\beta_R
+
2\Big(
-\sin t\,(\bar k\mu-A g-\beta_R)
+
(1-\cos t)\,\beta_I
\Big)
\Big].
\label{eq:app_opto_Fgt_full}
\end{equation}
Isolating the $g$-dependent term, $-\sin t\,(\bar k\mu-A g-\beta_R)=-\sin t\,(\bar k\mu-\beta_R)+A g\sin t$,
we obtain
\begin{equation}
F_{gt}(g,t)=d_0(t)+d_1(t)\,g.
\end{equation}
with
\begin{align}
d_0(t)
&=
A\Big[
-16\,\bar k^2\mu\,\zeta(t)\,(1-\cos t)\,\beta_R
+
2\Big(
-\sin t\,(\bar k\mu-\beta_R)
+
(1-\cos t)\,\beta_I
\Big)
\Big],
\\
d_1(t)
&=
2A^2\sin t.
\end{align}

This is the only structural fact required from the quoted benchmark expression in the main text: the closed optomechanical model belongs to the same affine-retention class as the free-fall and differential KC benchmarks. The derivation of the full table-row formulas for $g_c^{\rm om}$, $g_*^{\rm om}$, and $(\alpha_0^{\rm om},\alpha_1^{\rm om})$ is not contained in the present appendix text.

\section{Kasevich--Chu operator algebra and full-state benchmark derivations}
\label{app:kc_operator_algebra}

This appendix supplies the operator derivations used in Sec.~\ref{sec:discrete_KC}. We first derive the internal fringe phase for the standard three-pulse Kasevich--Chu (KC) sequence. We then establish the product-state identities needed for the full-state benchmark, including the motional contribution to the timing QFI and the cancelation of mixed motional terms in the gravity--time cross element.

\subsection{Internal KC fringe from pulse algebra}
\label{app:kc_internal_fringe}

We consider the symmetric $\pi/2$--$\pi$--$\pi/2$ sequence at pulse times
\begin{equation}
t_1=0,\qquad t_2=T,\qquad t_3=2T.
\label{eq:app_kc_times}
\end{equation}
The motional Hilbert space is denoted by $\mathcal H_{\rm mot}$ and the internal basis by $\{\ket{a},\ket{b}\}$. The total Hilbert space is $\mathcal H_{\rm mot}\otimes \mathbb C^2$. Define the phase operators
\begin{equation}
\Theta_j:=k_0\hat z(t_j)-\phi_j,
\qquad
E_j:=e^{+i\Theta_j},
\qquad
E_j^\dagger=e^{-i\Theta_j},
\end{equation}
where $\hat z(t_j)$ is the Heisenberg position operator evaluated at the $j$th pulse time.

For pulse area $\pi/2$, the internal pulse matrix is
\begin{equation} 
U_{\pi/2}(\Theta_j)
=
\frac{1}{\sqrt2}
\begin{pmatrix}
1 & -iE_j^\dagger\\
-iE_j & 1
\end{pmatrix}.
\label{eq:app_Upi2_matrix}
\end{equation}
and for pulse area $\pi$,
\begin{equation}
U_{\pi}(\Theta_j)
=
-i
\begin{pmatrix}
0 & E_j^\dagger\\
E_j & 0
\end{pmatrix}.
\label{eq:app_Upi_matrix}
\end{equation}
The total internal pulse sequence is therefore
\begin{equation} 
U_{\rm KC}=U_{\pi/2}(\Theta_3)\,U_{\pi}(\Theta_2)\,U_{\pi/2}(\Theta_1).
\label{eq:app_UKC_sequence}
\end{equation}

\subsubsection{Exact transition amplitude}

Let $U_1:=U_{\pi/2}(\Theta_1)$, $U_2:=U_{\pi}(\Theta_2)$, and $U_3:=U_{\pi/2}(\Theta_3)$. 
We first compute the product $U_2U_1$:
\begin{equation}
\begin{aligned}
U_2U_1
&=
\left(-i
\begin{pmatrix}
0 & E_2^\dagger\\
E_2 & 0
\end{pmatrix}
\right)
\left(
\frac{1}{\sqrt2}
\begin{pmatrix}
1 & -iE_1^\dagger\\
-iE_1 & 1
\end{pmatrix}
\right)
\\
&=
\frac{-i}{\sqrt2}
\begin{pmatrix}
0\cdot 1 + E_2^\dagger(-iE_1)
&
0\cdot(-iE_1^\dagger)+E_2^\dagger\cdot 1
\\[0.4em]
E_2\cdot 1+0\cdot(-iE_1)
&
E_2(-iE_1^\dagger)+0\cdot 1
\end{pmatrix}
\\
&=
\frac{-i}{\sqrt2}
\begin{pmatrix}
-iE_2^\dagger E_1 & E_2^\dagger\\
E_2 & -iE_2E_1^\dagger
\end{pmatrix}
=
\frac{1}{\sqrt2}
\begin{pmatrix}
-\,E_2^\dagger E_1 & -iE_2^\dagger\\
-iE_2 & -\,E_2E_1^\dagger
\end{pmatrix}.
\end{aligned}
\end{equation}

Now multiply by $U_3$ from the left. The amplitude to exit in $\ket{b}$ starting from $\ket{a}$ is the $(b,a)$ entry of $U_3U_2U_1$. Since row $b$ of $U_3$ is
\begin{equation} 
\frac{1}{\sqrt2}\,(-iE_3,\ 1),
\label{eq:app_U3_rowb}
\end{equation}
and column $a$ of $U_2U_1$ is
\begin{equation} 
\frac{1}{\sqrt2}
\binom{-E_2^\dagger E_1}{-iE_2},
\label{eq:app_U2U1_cola}
\end{equation}
their product gives
\begin{equation} 
\begin{aligned}
\hat A_b
&:=(U_3U_2U_1)_{ba}
=
\frac{1}{2}
\Big[
(-iE_3)(-E_2^\dagger E_1)+1\cdot(-iE_2)
\Big]
\\
&=
\frac{i}{2}\Big(E_3E_2^\dagger E_1-E_2\Big).
\end{aligned}
\label{eq:app_KC_Ab}
\end{equation}
This is the exact transition operator quoted in the main text.

\subsubsection{Population fringe and phase reduction}

The internal output probability is $P_b=\bra{\psi_0}\hat A_b^\dagger \hat A_b\ket{\psi_0}$. 
From Eq.~\eqref{eq:app_KC_Ab}, $\hat A_b^\dagger=-(i/2)\big(E_1^\dagger E_2E_3^\dagger-E_2^\dagger\big)$, 
so
\begin{equation} 
\begin{aligned}
\hat A_b^\dagger \hat A_b
&=
\frac{1}{4}
\Big(E_1^\dagger E_2E_3^\dagger-E_2^\dagger\Big)
\Big(E_3E_2^\dagger E_1-E_2\Big)
\\
&=
\frac{1}{4}
\Big(
E_1^\dagger E_2E_3^\dagger E_3E_2^\dagger E_1
-
E_1^\dagger E_2E_3^\dagger E_2
-
E_2^\dagger E_3E_2^\dagger E_1
+
E_2^\dagger E_2
\Big).
\end{aligned}
\end{equation}
Using $E_j^\dagger E_j=\mathbb I$, this becomes
\begin{equation} 
\hat A_b^\dagger \hat A_b
=
\frac12\,\mathbb I
-
\frac14
\Big(
E_1^\dagger E_2E_3^\dagger E_2
+
E_2^\dagger E_3E_2^\dagger E_1
\Big).
\label{eq:app_AbdagAb_reduced}
\end{equation}
The two nontrivial terms are Hermitian conjugates of one another, hence
\begin{equation}
P_b
=
\frac12
-
\frac12\,
\Re\,
\bra{\psi_0}
E_1^\dagger E_2E_3^\dagger E_2
\ket{\psi_0}.
\end{equation}

We now reduce the phase operator in the expectation value. Since each $\Theta_j$ is linear in $\hat z(t_j)$, all pairwise commutators are $c$-numbers. For such operators, the Weyl identity gives
\begin{equation}
e^{A_1}e^{A_2}\cdots e^{A_n}
=
\exp\!\left(
\sum_{r=1}^n A_r+\frac12\sum_{r<s}[A_r,A_s]
\right).
\label{eq:app_Weyl_identity}
\end{equation}
Applying this to
\begin{equation} 
E_1^\dagger E_2E_3^\dagger E_2
=
e^{-i\Theta_1}e^{+i\Theta_2}e^{-i\Theta_3}e^{+i\Theta_2},
\label{eq:app_E_product_chain}
\end{equation}
the exponent is $-i(\Theta_1-2\Theta_2+\Theta_3)$ 
plus a commutator correction. For the symmetric pulse times in Eq.~\eqref{eq:app_kc_times}, that correction cancels exactly, so
$E_1^\dagger E_2E_3^\dagger E_2=\exp\!\big[-i(\Theta_1-2\Theta_2+\Theta_3)\big]$. 

Under free fall in a uniform field, $\hat z(t)=\hat z+\hat p\,t/m-gt^2/2$. 
Therefore,
\begin{equation} 
\begin{aligned}
\hat z(0)-2\hat z(T)+\hat z(2T)
&=
\hat z
-2\left(\hat z+\frac{\hat p}{m}T-\frac12 gT^2\right)
+\left(\hat z+\frac{2\hat p}{m}T-2gT^2\right)
\\
&=
-gT^2\,\mathbb I.
\end{aligned}
\end{equation}
Substituting into the phase combination gives
\begin{equation}
\Theta_1-2\Theta_2+\Theta_3
=
-k_0gT^2-(\phi_1-2\phi_2+\phi_3)
=: \Delta\Phi(g,T).
\label{eq:app_DeltaPhi}
\end{equation}
Hence $E_1^\dagger E_2E_3^\dagger E_2=e^{-i\Delta\Phi(g,T)}$, 
and therefore
\begin{equation}
P_b
=
\frac12\Big[1-C\cos\Delta\Phi(g,T)\Big].
\end{equation}
where $C$ denotes the fringe contrast. Under ideal closure one has $C=1$.

\subsection{Differential KC benchmark: additive QFI and motional timing contribution}
\label{app:kc_full_state}

We now derive the identities used in the full-state benchmark of Sec.~\ref{subsec:KC_full_state}. Under ideal closure, the final state can be written as
\begin{equation}
\ket{\Psi(g,T)}
=
\ket{\psi_{\rm out}(T)}\otimes\ket{\chi(g,T)},
\end{equation}
where $\ket{\psi_{\rm out}(T)}$ is a motional state independent of the differential gravity phase and
\begin{equation}
\ket{\chi(g,T)}
=
\frac{1}{\sqrt2}
\left(
\ket{a}+e^{i\Delta\Phi(g,T)}\ket{b}
\right).
\end{equation}

\subsubsection{Additivity of the timing QFI}

For a product pure state $\ket{\Psi(T)}=\ket{\psi(T)}\otimes\ket{\chi(T)}$,
the derivative is $\partial_T\ket{\Psi}=\ket{\psi_T}\otimes\ket{\chi}+\ket{\psi}\otimes\ket{\chi_T}$,
where $\ket{\psi_T}:=\partial_T\ket{\psi(T)}$ and $\ket{\chi_T}:=\partial_T\ket{\chi(T)}$. 
Therefore
\begin{align}
\braket{\partial_T\Psi|\partial_T\Psi}
&=
\braket{\psi_T|\psi_T}
+
\braket{\chi_T|\chi_T}
+
\braket{\psi_T|\psi}\braket{\chi|\chi_T}
+
\braket{\psi|\psi_T}\braket{\chi_T|\chi},
\\
\braket{\Psi|\partial_T\Psi}
&=
\braket{\psi|\psi_T}
+
\braket{\chi|\chi_T}.
\end{align}
Taking the modulus squared of the second expression gives
\begin{align}
\big|\braket{\Psi|\partial_T\Psi}\big|^2
&=
\big|\braket{\psi|\psi_T}\big|^2
+
\big|\braket{\chi|\chi_T}\big|^2
+
\braket{\psi_T|\psi}\braket{\chi|\chi_T}
+
\braket{\psi|\psi_T}\braket{\chi_T|\chi}.
\end{align}
The mixed terms cancel in the pure-state QFI combination $4\big(\braket{\partial_T\Psi|\partial_T\Psi}-|\braket{\Psi|\partial_T\Psi}|^2\big)$,
hence
\begin{equation}
F_{TT}^{(\mathrm{diff})}
=
F_{TT}^{(\mathrm{mot})}
+
F_{TT}^{(\mathrm{int})}.
\label{eq:app_QFI_additivity}
\end{equation}

\subsubsection{Motional timing generator}

The motional factor may be written as
\begin{equation}
\ket{\psi_{\rm out}(T)}
=
\hat U_{\rm mot}(T)\ket{\psi_0},
\qquad
\hat U_{\rm mot}(T)
=
\exp\!\left[i k_0\left(\hat z+\frac{\hat p}{m}T\right)\right].
\end{equation}
Let
\begin{equation}
\hat K(T):=
ik_0\hat z+i\frac{k_0}{m}T\hat p,
\qquad
\hat U_{\rm mot}(T)=e^{\hat K(T)}.
\end{equation}
The pulled-back local timing generator is
\begin{equation}
\hat{\mathcal H}^{(\mathrm{mot})}_T
=
i\,\hat U_{\rm mot}^\dagger(T)\,\partial_T\hat U_{\rm mot}(T).
\end{equation}
Using the integral identity
\begin{equation}
\partial_T e^{\hat K(T)}
=
\int_0^1 e^{(1-s)\hat K(T)}\big(\partial_T\hat K(T)\big)e^{s\hat K(T)}\,ds,
\label{eq:app_dexp_identity}
\end{equation}
we find
\begin{equation}
\hat{\mathcal H}^{(\mathrm{mot})}_T
=
i\int_0^1 e^{-s\hat K(T)}
\big(\partial_T\hat K(T)\big)
e^{s\hat K(T)}\,ds
=
-\frac{k_0}{m}\int_0^1 e^{-s\hat K(T)}\hat p\,e^{s\hat K(T)}\,ds.
\end{equation}
Since $[\hat p,\hat K(T)]=k_0\hbar$,
all higher nested commutators vanish, and therefore
$e^{-s\hat K(T)}\hat p\,e^{s\hat K(T)}=\hat p+s\,k_0\hbar$.
Integrating over $s$ gives
\begin{equation}
\hat{\mathcal H}^{(\mathrm{mot})}_T
=
-\frac{k_0}{m}
\left(
\hat p+\frac{k_0\hbar}{2}
\right).
\end{equation}
The additive scalar has no effect on the variance, hence
\begin{equation}
F_{TT}^{(\mathrm{mot})}
=
4\,\operatorname{Var}_{\psi_0}\!\left(\hat{\mathcal H}^{(\mathrm{mot})}_T\right)
=
\frac{4k_0^2}{m^2}\operatorname{Var}_{\psi_0}(\hat p).
\end{equation}
This is the motional timing contribution quoted in the main text.

\subsubsection{Cross-term cancellation and internal phase dependence}

We now derive the gravity-time cross element for the factorized full state. Since the motional factor is independent of the differential gravity phase,
\begin{equation}
\partial_g\ket{\Psi}
=
\ket{\psi}\otimes\ket{\chi_g},
\qquad
\partial_T\ket{\Psi}
=
\ket{\psi_T}\otimes\ket{\chi}
+
\ket{\psi}\otimes\ket{\chi_T},
\end{equation}
where $\ket{\chi_g}:=\partial_g\ket{\chi(g,T)}$. 
Therefore
\begin{align}
\braket{\partial_g\Psi|\partial_T\Psi}
&=
\braket{\psi|\psi_T}\braket{\chi_g|\chi}
+
\braket{\chi_g|\chi_T},
\\
\braket{\partial_g\Psi|\Psi}
&=
\braket{\chi_g|\chi},
\\
\braket{\Psi|\partial_T\Psi}
&=
\braket{\psi|\psi_T}
+
\braket{\chi|\chi_T}.
\end{align}
Substituting these expressions into the pure-state QFIM formula,
\begin{equation}
F_{gT}^{(\mathrm{diff})}
=
4\,\Re\!\left(
\braket{\partial_g\Psi|\partial_T\Psi}
-
\braket{\partial_g\Psi|\Psi}\braket{\Psi|\partial_T\Psi}
\right),
\label{eq:app_cross_QFIM_def}
\end{equation}
the mixed motional term $\braket{\psi|\psi_T}$ cancels exactly, leaving
\begin{equation}
F_{gT}^{(\mathrm{diff})}
=
4\,\Re\!\left(
\braket{\chi_g|\chi_T}
-
\braket{\chi_g|\chi}\braket{\chi|\chi_T}
\right).
\label{eq:app_cross_reduced}
\end{equation}
Thus the full-state gravity-time coupling is determined entirely by the internal phase qubit.

For $\ket{\chi(g,T)}=\big(\ket{a}+e^{i\Delta\Phi(g,T)}\ket{b}\big)/\sqrt2$, 
one has $\partial_j\ket{\chi}=i\big(\partial_j\Delta\Phi\big)e^{i\Delta\Phi}\ket{b}/\sqrt2$, $j\in\{g,T\}$. 
A direct substitution into Eq.~\eqref{eq:app_cross_reduced} gives
\begin{equation}
F_{gT}^{(\mathrm{diff})}
=
(\partial_g\Delta\Phi)(\partial_T\Delta\Phi).
\end{equation}
Using $\Delta\Phi(g,T)=-k_0gT^2-\phi_{\rm ctrl}$,
we obtain
\begin{equation}
F_{gT}^{(\mathrm{diff})}(g,T)
=
(-k_0T^2)(-2k_0gT)
=
2k_0^2gT^3.
\end{equation}
This is the cross element quoted in the main text.
\end{document}